\pdfoutput=1
\documentclass[12pt,a4paper]{article}
\usepackage{ifthen}
\newboolean{pdflatex}
\setboolean{pdflatex}{true}

\newboolean{articletitles}
\setboolean{articletitles}{true}

\newboolean{uprightparticles}
\setboolean{uprightparticles}{false}

\newboolean{inbibliography}
\setboolean{inbibliography}{false}

\def\paperauthors{LHCb collaboration} 
\def\paperasciititle{Search for the rare decays B^0_s->e^+e^- and B^0->e^+e^-}  
\def\papertitle{Search for the rare decays $B^0_s\to e^+e^-$ and $B^0\to e^+e^-$}
\def\paperkeywords{{High Energy Physics}, {LHCb}, {Rare Decay},{Flavour Physics}, {B physics}, {Branching Fraction}}
\def\papercopyright{2020 CERN for the benefit of the LHCb collaboration}
\def\paperlicence{CC BY 4.0 licence}
\def\paperlicenceurl{https://creativecommons.org/licenses/by/4.0/}

\usepackage[top=1in, bottom=1.25in, left=1in, right=1in]{geometry}

\usepackage{microtype}
\usepackage{lineno}  
\usepackage{xspace} 
\usepackage{caption} 

\usepackage{graphicx}  
\usepackage{color}
\usepackage{colortbl}
\graphicspath{{./figs/}} 

\usepackage{amsmath} 
\usepackage{amssymb}
\usepackage{amsfonts}
\usepackage{upgreek} 

\newcommand*\patchAmsMathEnvironmentForLineno[1]{%
\expandafter\let\csname old#1\expandafter\endcsname\csname #1\endcsname
\expandafter\let\csname oldend#1\expandafter\endcsname\csname
end#1\endcsname
 \renewenvironment{#1}%
   {\linenomath\csname old#1\endcsname}%
   {\csname oldend#1\endcsname\endlinenomath}%
}
\newcommand*\patchBothAmsMathEnvironmentsForLineno[1]{%
  \patchAmsMathEnvironmentForLineno{#1}%
  \patchAmsMathEnvironmentForLineno{#1*}%
}
\AtBeginDocument{%
\patchBothAmsMathEnvironmentsForLineno{equation}%
\patchBothAmsMathEnvironmentsForLineno{align}%
\patchBothAmsMathEnvironmentsForLineno{flalign}%
\patchBothAmsMathEnvironmentsForLineno{alignat}%
\patchBothAmsMathEnvironmentsForLineno{gather}%
\patchBothAmsMathEnvironmentsForLineno{multline}%
\patchBothAmsMathEnvironmentsForLineno{eqnarray}%
}

\usepackage{hyperxmp}

\usepackage[pdftex,
            pdfauthor={\paperauthors},
            pdftitle={\paperasciititle},
            pdfkeywords={\paperkeywords},
            pdfcopyright={Copyright (C) \papercopyright},
            pdflicenseurl={\paperlicenceurl}]{hyperref}

\usepackage[colorinlistoftodos,textsize=scriptsize]{todonotes}

\usepackage[bottom,flushmargin,hang,multiple]{footmisc}

\usepackage[all]{hypcap} 

\usepackage{xspace} 
\usepackage{upgreek}

\def\lhcb   {\mbox{LHCb}\xspace}

\def\MagUp {\mbox{\em Mag\kern -0.05em Up}\xspace}

\ifthenelse{\boolean{uprightparticles}}%
{

 \def\Pmu         {\ensuremath{\upmu}\xspace}                 
 \def\Pnu         {\ensuremath{\upnu}\xspace}                 
                  
 \def\Ppi         {\ensuremath{\uppi}\xspace}

 \def\Ppsi        {\ensuremath{\uppsi}\xspace}

 \def\PDelta      {\ensuremath{\Delta}\xspace}                 
 \def\PXi         {\ensuremath{\Xi}\xspace}                 
 \def\PLambda     {\ensuremath{\Lambda}\xspace}                 
 \def\PSigma      {\ensuremath{\Sigma}\xspace}                 
 \def\POmega      {\ensuremath{\Omega}\xspace}                 
 \def\PUpsilon    {\ensuremath{\Upsilon}\xspace}

 \def\PB      {\ensuremath{\mathrm{B}}\xspace}                 
                  
 \def\PD      {\ensuremath{\mathrm{D}}\xspace}

 \def\PJ      {\ensuremath{\mathrm{J}}\xspace}                 
 \def\PK      {\ensuremath{\mathrm{K}}\xspace}

 \def\Pb      {\ensuremath{\mathrm{b}}\xspace}                 
 \def\Pc      {\ensuremath{\mathrm{c}}\xspace}                 
                  
 \def\Pe      {\ensuremath{\mathrm{e}}\xspace}

 \def\Ph      {\ensuremath{\mathrm{h}}\xspace}                 
 \def\Pi      {\ensuremath{\mathrm{i}}\xspace}

 \def\Pp      {\ensuremath{\mathrm{p}}\xspace}

 \def\Ps      {\ensuremath{\mathrm{s}}\xspace}

 \def\thebaroffset{0.0em}
}
{

 \def\Pmu         {\ensuremath{\mu}\xspace}                 
 \def\Pnu         {\ensuremath{\nu}\xspace}                 
                  
 \def\Ppi         {\ensuremath{\pi}\xspace}

 \def\Ppsi        {\ensuremath{\psi}\xspace}                 
                  
 \mathchardef\PDelta="7101
 \mathchardef\PXi="7104
 \mathchardef\PLambda="7103
 \mathchardef\PSigma="7106
 \mathchardef\POmega="710A
 \mathchardef\PUpsilon="7107
                  
 \def\PB      {\ensuremath{B}\xspace}                 
                  
 \def\PD      {\ensuremath{D}\xspace}

 \def\PJ      {\ensuremath{J}\xspace}                 
 \def\PK      {\ensuremath{K}\xspace}

 \def\Pb      {\ensuremath{b}\xspace}                 
 \def\Pc      {\ensuremath{c}\xspace}                 
                  
 \def\Pe      {\ensuremath{e}\xspace}

 \def\Ph      {\ensuremath{h}\xspace}                 
 \def\Pi      {\ensuremath{i}\xspace}

 \def\Pp      {\ensuremath{p}\xspace}

 \def\Ps      {\ensuremath{s}\xspace}

 \def\thebaroffset{0.18em}
}
\newcommand{\offsetoverline}[2][\thebaroffset]{\kern #1\overline{\kern -#1 #2}}%

\makeatletter
\ifcase \@ptsize \relax
  \newcommand{\miniscule}{\@setfontsize\miniscule{4}{5}}
\or
  \newcommand{\miniscule}{\@setfontsize\miniscule{5}{6}}
\or
  \newcommand{\miniscule}{\@setfontsize\miniscule{5}{6}}
\fi
\makeatother

\DeclareRobustCommand{\optbar}[1]{\shortstack{{\miniscule (\rule[.5ex]{1.25em}{.18mm})}
  \\ [-.7ex] $#1$}}

\def\en         {{\ensuremath{\Pe^-}}\xspace}   
\def\ep         {{\ensuremath{\Pe^+}}\xspace}

\def\epem       {{\ensuremath{\Pe^+\Pe^-}}\xspace}

\def\mup        {{\ensuremath{\Pmu^+}}\xspace}
\def\mun        {{\ensuremath{\Pmu^-}}\xspace} 

\def\mumu       {{\ensuremath{\Pmu^+\Pmu^-}}\xspace}

\def\ellell     {\ensuremath{\ell^+ \ell^-}\xspace}

\def\neu        {{\ensuremath{\Pnu}}\xspace}
\def\neub       {{\ensuremath{\overline{\Pnu}}}\xspace}
\def\neue       {{\ensuremath{\neu_e}}\xspace}
\def\neueb      {{\ensuremath{\neub_e}}\xspace}

\def\squark    {{\ensuremath{\Ps}}\xspace}

\def\cquark    {{\ensuremath{\Pc}}\xspace}

\def\bquark    {{\ensuremath{\Pb}}\xspace}

\def\hadron {{\ensuremath{\Ph}}\xspace}
\def\pion   {{\ensuremath{\Ppi}}\xspace}

\def\pip    {{\ensuremath{\pion^+}}\xspace}
\def\pim    {{\ensuremath{\pion^-}}\xspace}

\def\kaon    {{\ensuremath{\PK}}\xspace}

\def\KorKbar {\kern \thebaroffset\optbar{\kern -\thebaroffset \PK}{}\xspace}

\def\Kp      {{\ensuremath{\kaon^+}}\xspace}
\def\Km      {{\ensuremath{\kaon^-}}\xspace}

\def\Kstarz  {{\ensuremath{\kaon^{*0}}}\xspace}

\def\Dbar    {{\ensuremath{\offsetoverline{\PD}}}\xspace}
\def\D       {{\ensuremath{\PD}}\xspace}

\def\DorDbar {\kern \thebaroffset\optbar{\kern -\thebaroffset \PD}\xspace}
\def\Dz      {{\ensuremath{\D^0}}\xspace}
\def\Dzb     {{\ensuremath{\Dbar{}^0}}\xspace}
\def\Dp      {{\ensuremath{\D^+}}\xspace}
\def\Dm      {{\ensuremath{\D^-}}\xspace}

\def\DpDm    {\ensuremath{\Dp {\kern -0.16em \Dm}}\xspace}

\def\B       {{\ensuremath{\PB}}\xspace}

\def\BorBbar {\kern \thebaroffset\optbar{\kern -\thebaroffset \PB}\xspace}
\def\Bz      {{\ensuremath{\B^0}}\xspace}

\def\Bd      {{\ensuremath{\B^0}}\xspace}

\def\BdorBdbar {\kern \thebaroffset\optbar{\kern -\thebaroffset \Bd}\xspace}
\def\Bu      {{\ensuremath{\B^+}}\xspace}

\def\Bp      {{\ensuremath{\Bu}}\xspace}

\def\Bs      {{\ensuremath{\B^0_\squark}}\xspace}

\def\BsorBsbar {\kern \thebaroffset\optbar{\kern -\thebaroffset \Bs}\xspace}

\def\Bds     {{\ensuremath{\B_{(\squark)}^0}}\xspace}

\def\BdorBs  {\Bds\xspace}

\def\jpsi     {{\ensuremath{{\PJ\mskip -3mu/\mskip -2mu\Ppsi\mskip 2mu}}}\xspace}

\def\Y#1S{\ensuremath{\PUpsilon{(#1S)}}\xspace}

\def\proton      {{\ensuremath{\Pp}}\xspace}

\def\Lz          {{\ensuremath{\PLambda}}\xspace}

\def\LorLbar     {\kern \thebaroffset\optbar{\kern -\thebaroffset \PLambda}\xspace}

\def\Lb           {{\ensuremath{\Lz^0_\bquark}}\xspace}

\newcommand{\decay}[2]{\ensuremath{#1\!\to #2}\xspace} 

\def\to                 {\ensuremath{\rightarrow}\xspace}

\def\CP                {{\ensuremath{C\!P}}\xspace}

\def\BsToJPsiPhi  {\decay{\Bs}{\jpsi\phi}}

\def\AT#1     {\ensuremath{A_{\mathrm{T}}^{#1}}\xspace}           

\def\Bsmm     {\decay{\Bs}{\mup\mun}}

\def\Bsee     {\decay{\Bs}{\epem}}
\def\Bdee     {\decay{\Bd}{\epem}}

\def\C#1      {\ensuremath{\mathcal{C}_{#1}}\xspace}                       
\def\Cp#1     {\ensuremath{\mathcal{C}_{#1}^{'}}\xspace}                    
\def\Ceff#1   {\ensuremath{\mathcal{C}_{#1}^{\mathrm{(eff)}}}\xspace}        
\def\Cpeff#1  {\ensuremath{\mathcal{C}_{#1}^{'\mathrm{(eff)}}}\xspace}       
\def\Ope#1    {\ensuremath{\mathcal{O}_{#1}}\xspace}                       
\def\Opep#1   {\ensuremath{\mathcal{O}_{#1}^{'}}\xspace}                    

\newcommand{\aunit}[1]{\ensuremath{\text{\,#1}}}       

\newcommand{\tev}{\aunit{Te\kern -0.1em V}\xspace}
\newcommand{\gev}{\aunit{Ge\kern -0.1em V}\xspace}
\newcommand{\mev}{\aunit{Me\kern -0.1em V}\xspace}
\newcommand{\kev}{\aunit{ke\kern -0.1em V}\xspace}
\newcommand{\ev}{\aunit{e\kern -0.1em V}\xspace}
\newcommand{\mevc}{\ensuremath{\aunit{Me\kern -0.1em V\!/}c}\xspace}
\newcommand{\gevc}{\ensuremath{\aunit{Ge\kern -0.1em V\!/}c}\xspace}
\newcommand{\mevcc}{\ensuremath{\aunit{Me\kern -0.1em V\!/}c^2}\xspace}
\newcommand{\gevcc}{\ensuremath{\aunit{Ge\kern -0.1em V\!/}c^2}\xspace}

\def\fb   {\ensuremath{\aunit{fb}}\xspace}
\def\invfb   {\ensuremath{\fb^{-1}}\xspace}

\def\gsim{{~\raise.15em\hbox{$>$}\kern-.85em
          \lower.35em\hbox{$\sim$}~}\xspace}
\def\lsim{{~\raise.15em\hbox{$<$}\kern-.85em
          \lower.35em\hbox{$\sim$}~}\xspace}

\def\pt         {\ensuremath{p_{\mathrm{T}}}\xspace}

\def\evtgen     {\mbox{\textsc{EvtGen}}\xspace}

\def\geant      {\mbox{\textsc{Geant4}}\xspace}

\def\photos     {\mbox{\textsc{Photos}}\xspace}

\def\pythia     {\mbox{\textsc{Pythia}}\xspace}

\def\tell1  {TELL1\xspace}
\def\ukl1   {UKL1\xspace}

\usepackage{cite} 
\usepackage{mciteplus}

\usepackage{booktabs}
\usepackage[separate-uncertainty=true]{siunitx}

\def\BToJpsiK {\ensuremath{\decay{\Bu}{\jpsi\Kp}}\xspace}
\def\BsToJPsiPhi  {\ensuremath{\decay{\Bs}{\jpsi\phi}}\xspace}
\def\Bee {\decay{\BdorBs}{\epem}}

\begin{document}

\renewcommand{\thefootnote}{\fnsymbol{footnote}}
\setcounter{footnote}{1}

\begin{titlepage}
\pagenumbering{roman}

\vspace*{-1.5cm}
\centerline{\large EUROPEAN ORGANIZATION FOR NUCLEAR RESEARCH (CERN)}
\vspace*{1.5cm}
\noindent
\begin{tabular*}{\linewidth}{lc@{\extracolsep{\fill}}r@{\extracolsep{0pt}}}
\ifthenelse{\boolean{pdflatex}}%
{\vspace*{-1.5cm}\mbox{\!\!\!\includegraphics[width=.14\textwidth]{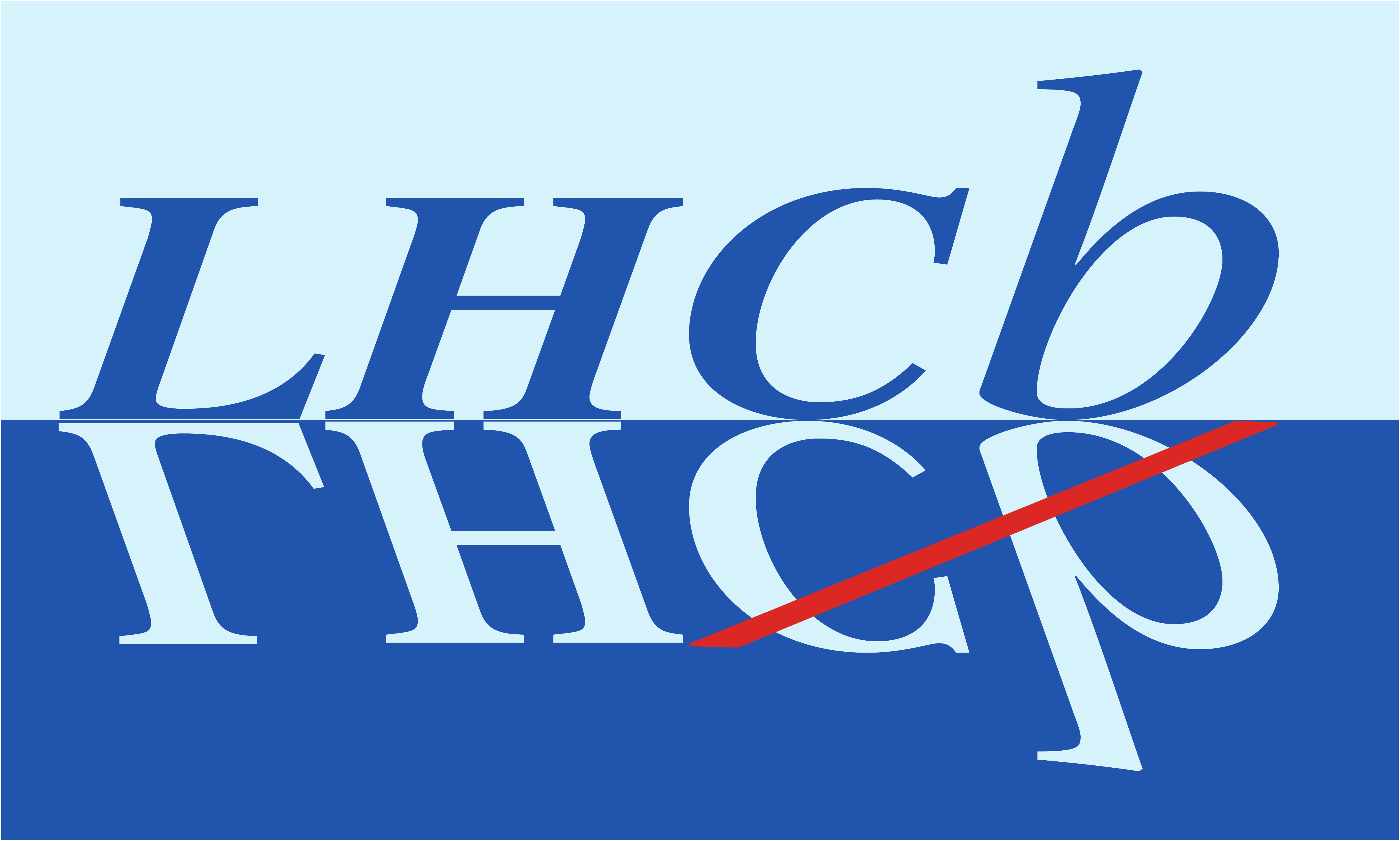}} & &}%
{\vspace*{-1.2cm}\mbox{\!\!\!\includegraphics[width=.12\textwidth]{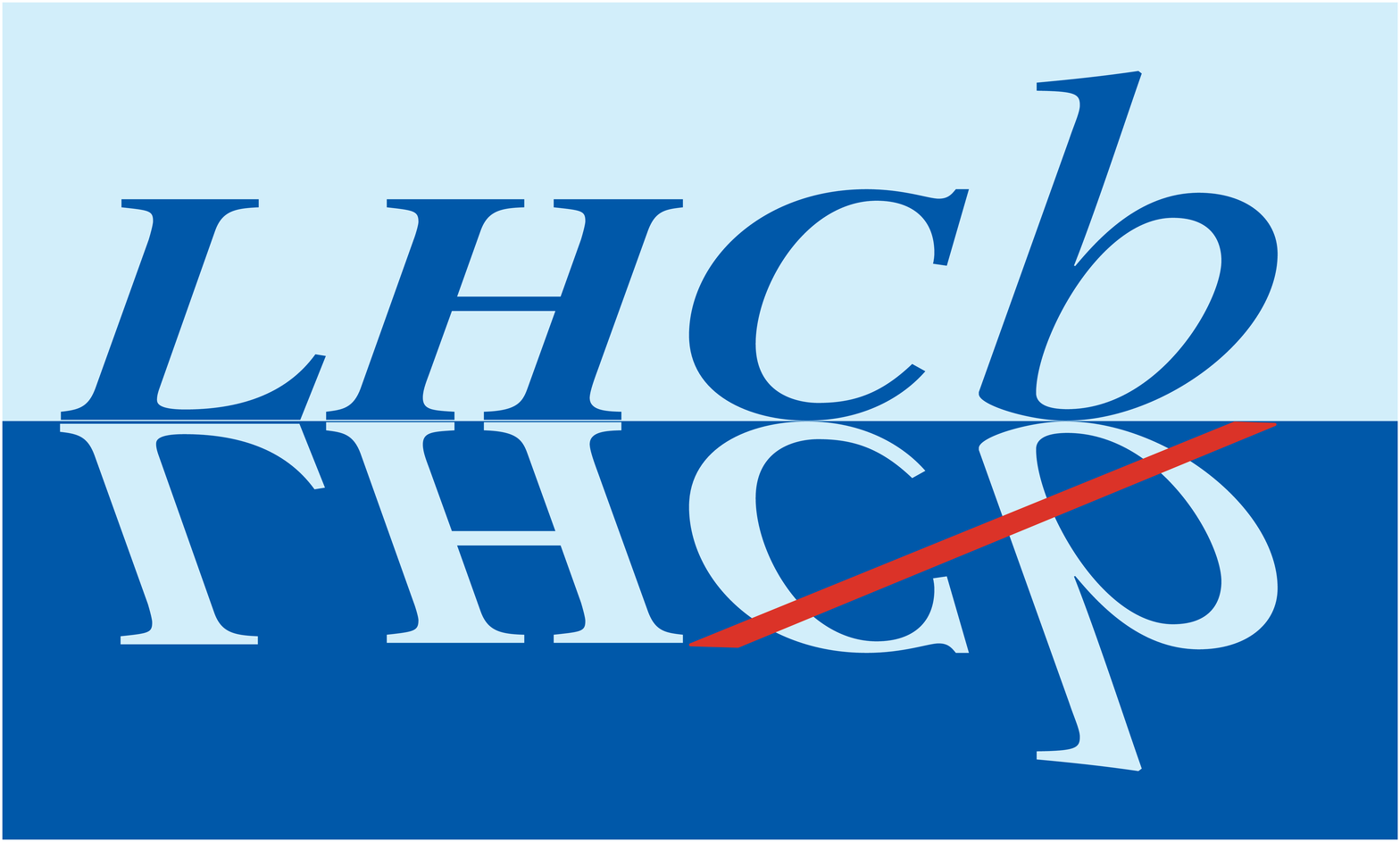}} & &}%
\\
 & & CERN-EP-2020-023 \\  
 & & LHCb-PAPER-2020-001 \\  
 & & May 28, 2020 \\ 
 & & \\
\end{tabular*}

\vspace*{4.0cm}

{\normalfont\bfseries\boldmath\huge
\begin{center}
  \papertitle 
\end{center}
}

\vspace*{2.0cm}

\begin{center}
\paperauthors\footnote{Authors are listed at the end of this Letter.}
\end{center}

\vspace{\fill}

\begin{abstract}
  \noindent
  A search for the decays $B^0_s\to e^+e^-$ and $B^0\to e^+e^-$ is performed using data collected with the LHCb experiment in proton-proton collisions at center-of-mass
energies of $7$, $8$ and $13\,\text{TeV}$, corresponding to integrated luminosities of $1$, $2$ and $2\,\text{fb}^{-1}$, respectively. No signal is observed. Assuming no contribution from $B^0\to e^+e^-$ decays, an upper limit on the branching fraction $\mathcal{B}(B^0_s\to e^+e^-)<9.4\,(11.2)\times10^{-9}$ is obtained at $90\,(95)\,\%$ confidence level. If no $B^0_s\to e^+e^-$ contribution is assumed, a limit of $\mathcal{B}(B^0\to e^+e^-)<2.5\,(3.0)\times10^{-9}$ is determined at $90\,(95)\,\%$ confidence level.
  These upper limits are more than one order of magnitude lower than the previous values.
\end{abstract}

\vspace*{2.0cm}

\begin{center}
  Published in Phys.~Rev.~Lett. 124 (2020) 211802
\end{center}

\vspace{\fill}

{\footnotesize 
\centerline{\copyright~\papercopyright. \href{\paperlicenceurl}{\paperlicence}.}}
\vspace*{2mm}

\end{titlepage}

\newpage
\setcounter{page}{2}
\mbox{~}

\renewcommand{\thefootnote}{\arabic{footnote}}
\setcounter{footnote}{0}

\cleardoublepage


\pagestyle{plain} 
\setcounter{page}{1}
\pagenumbering{arabic}

\noindent Searches for rare particle decays provide ideal probes for contributions from physics processes beyond the Standard Model (SM). 
Recent measurements of decays involving $\decay{b}{s\ell^+\ell^-}$ transitions (the inclusion of charge-conjugated processes is implied throughout this Letter) hint at deviations from SM predictions in lepton-flavor universality tests~\cite{LHCb-PAPER-2017-013,LHCb-PAPER-2019-009,LHCb-PAPER-2019-040,Lees:2012tva,Abdesselam:2019lab,Abdesselam:2019wac} and thus motivate measurements of decay rates into final states involving leptons. Following the observation of the decay \Bsmm~\cite{LHCB-PAPER-2014-049,LHCb-PAPER-2017-001}, the search for \Bsee and \Bdee decays provides an independent test of lepton-flavor universality.
According to SM predictions (calculated from Ref.~\cite{Beneke:2019slt}, neglecting QED corrections that are expected to be at the percent level), $\mbox{\Bee}$ decays have branching fractions of  \mbox{$\mathcal{B}(\Bsee)=\num{8.60\pm0.36e-14}$ } and \mbox{$\mathcal{B}(\Bdee)=\num{2.41\pm0.13e-15}$}.
With contributions beyond the SM, these branching fractions could be significantly larger, reaching values of $\mathcal{O}(10^{-8})$ for $\mathcal{B}(\Bsee)$ and $\mathcal{O}(10^{-10})$ for $\mathcal{B}\mbox{(\Bdee)}$~\cite{Fleischer:2017ltw}. These values are close to the current experimental bounds of  \mbox{$\mathcal{B}(\Bsee)<\num{2.8e-7}$} and \mbox{$\mathcal{B}(\Bdee)<\num{8.3e-8}$} at $\SI{90}{\percent}$ confidence level (CL)~\cite{Aaltonen:2009vr}, set by the CDF collaboration.

In this Letter a search for \Bsee and \Bdee decays is presented using data collected with the LHCb experiment in proton-proton collisions at center-of-mass energies of $\SI{7}{TeV}$ in 2011, $\SI{8}{TeV}$ in 2012 and $\SI{13}{TeV}$ in 2015 and 2016, corresponding to integrated luminosities of $1$, $2$ and $\SI{2}{fb^{-1}}$, respectively.
The signal yields are determined from a fit to the data and normalized to those of the \mbox{\BToJpsiK} decay, where the \jpsi meson decays to \epem, which has a precisely measured branching fraction~\cite{PDG2019} and a similar dielectron signature in the detector.

The \lhcb detector~\cite{LHCb-DP-2008-001,LHCb-DP-2014-002} is a single-arm forward spectrometer covering the \mbox{pseudorapidity} range $2< \eta <5$,
designed for the study of particles containing \bquark or \cquark
quarks. The detector includes a high-precision tracking system
consisting of a silicon-strip vertex detector surrounding the $pp$
interaction region, a large-area silicon-strip detector located
upstream of a dipole magnet with a bending power of about
$4{\mathrm{\,Tm}}$, and three stations of silicon-strip detectors and straw
drift tubes placed downstream of the magnet.
Different types of charged hadrons are distinguished using information
from two ring-imaging Cherenkov detectors. 
Photons, electrons and hadrons are identified by a calorimeter system consisting of
scintillating-pad and preshower detectors, an electromagnetic
and a hadronic calorimeter. Muons are identified by a
system composed of alternating layers of iron and multiwire
proportional chambers.

The online event selection is performed by a trigger~\cite{LHCb-DP-2012-004}, 
which consists of a hardware stage, based on information from the calorimeter and muon
systems, followed by a software stage, which applies a full event
reconstruction.
At the hardware trigger stage, events are required to have a high-energy deposit in the 
calorimeters associated with a signal electron candidate, or a muon candidate with high transverse 
momentum \pt, or a photon, electron or hadron candidate with high transverse energy from the decays
of other particles from the $pp$ collision.
The software trigger requires a two-track secondary vertex with a significant displacement from any primary
$pp$ interaction vertex (PV). At least one charged particle
must have high \pt and be inconsistent with originating from a PV.
A multivariate algorithm~\cite{BBDT,LHCb-PROC-2015-018} is used in the trigger for
the identification of secondary vertices consistent with the decay
of a \bquark hadron.
Simulated samples are used to optimize the candidate selection, estimate selection efficiencies and describe the expected invariant-mass shapes of the signal candidates and background decays.
  In the simulation, $pp$ collisions are generated using
\pythia~\cite{Sjostrand:2006za,*Sjostrand:2007gs} with a specific \lhcb
configuration~\cite{LHCb-PROC-2010-056}. Decays of unstable particles
are described by \evtgen~\cite{Lange:2001uf}, in which final-state
radiation is generated using \photos~\cite{Golonka:2005pn}. The
interaction of the generated particles with the detector, and its response,
are implemented using the \geant
toolkit~\cite{Allison:2006ve, *Agostinelli:2002hh} as described in
Ref.~\cite{LHCb-PROC-2011-006}. The simulation is corrected for data-simulation differences in \B-meson production kinematics, detector occupancy and isolation criteria~\cite{LHCb-PAPER-2011-025} using \BToJpsiK and \BsToJPsiPhi decays, with $\decay{\jpsi}{\epem}$ and $\decay{\phi}{\Kp\Km}$.
Particle identification variables are calibrated using data from \BToJpsiK and $\mbox{\decay{\Dz}{\Km\pip}}$ decays \cite{LHCb-PUB-2016-021}. 
The calibration data are binned in momentum and pseudorapidity of the particle as well as detector occupancy to account for possible differences in kinematics between the investigated decay and the calibration data.

The \Bee candidates are selected in events passing the trigger requirements by combining two tracks that are inconsistent with originating from any PV in the event and which form a good-quality secondary vertex.
The tracks are also required to have a momentum larger than 3\,\gevc and \pt greater than 500\,\mevc, and must be identified as electrons using information from the Cherenkov detectors and calorimeters.
The dielectron candidate's momentum must be aligned with the vector pointing from a PV (the associated PV) to the two-track vertex and have a considerable transverse component. The candidate must also have an invariant mass in the range $[4166,6566]\,\mevcc$.

The measured electron momenta are corrected
for losses due to bremsstrahlung radiation by adding the momentum of photons consistent with being
emitted upstream of the magnet \cite{LHCb-PAPER-2013-005}.
Candidates in data and simulation are separated into three categories with either zero, one, or both electrons having a bremsstrahlung correction applied.
To avoid experimenters' bias, the narrowest dielectron invariant-mass region containing $\SI{90}{\percent}$ of simulated \Bsee decays, corresponding to a range of [$4689$, $5588$]\,\mevcc, was removed from the data set until the analysis procedure was finalized.

Candidates for the normalization mode, \BToJpsiK, are constructed similarly, but require an additional track consistent with being a kaon and originating from the same vertex as the dielectron candidate.
The dielectron candidate must have an invariant mass in the range $[2450,3176]\,\mevcc$, consistent with arising from a \jpsi meson decay. In addition, the reconstructed \Bu candidate mass, when the dielectron candidate is constrained to the known \jpsi mass \cite{PDG2019}, must be above $5175\,\mevcc$, suppressing partially reconstructed decays.

A boosted decision tree (BDT) algorithm\cite{AdaBoost,Breiman,Scikit-learn-paper} is used to separate \Bee signal from random combinations of two electrons (combinatorial background). The BDT is trained separately for data taking periods 2011--2012 $\mbox{\text{(Run 1)}}$ and 2015--2016 $\mbox{\text{(Run 2)}}$ on simulated \Bsee decays as signal proxy and dielectron candidates from data with a mass above $5588\,\mevcc$ as background proxy.
The split between the data taking periods is done to account for changes in the center-of-mass energies and trigger strategies, which significantly impact the data distributions and improve the BDT and the particle identification algorithms in Run 2.
It is checked that the data behave consistently across the data-taking periods.
The BDT input variables comprise of the following: kinematic information on the electron tracks and \B candidate, information on the displacement of the electrons and \B candidate from the associated PV, and isolation variables that quantify the compatibility of other tracks in the event with originating from the same decay as the \B candidate \cite{LHCb-PAPER-2011-025,Gavardi:1645251}.
Candidates with a BDT response compatible with that of the background are discarded, with the threshold chosen by maximizing the figure of merit  ${\epsilon_{\text{signal}}}/{(\sqrt{N_{\text{background}}} + 3/2)}$~\cite{Punzi:2003bu},
where $\epsilon_{\text{signal}}$ is the signal efficiency and the expected background yield in the signal region is  $N_{\text{background}}$.

The final selected data set is separated by data-taking period and by category of bremsstrahlung correction.
The branching fraction $\mathcal{B}(\Bee)$ is measured relative to that of the normalization channel via
\begin{align}
\label{eqn:BRalpha}
	\mathcal{B}(\Bee) & =   N(\Bee)  \times \alpha \times \mathcal{B}(\BToJpsiK)\times \left(\frac{f_{d(s)}}{f_u}\right)^{-1},
\end{align}
where
\begin{align}
    \alpha& \equiv \frac{\varepsilon(\BToJpsiK)}{\varepsilon(\Bee)}\times \frac{1}{N(\BToJpsiK)},
\end{align}
$\varepsilon(\Bee)$ and $\varepsilon(\BToJpsiK)$ denote the efficiencies of the signal and normalization modes, and \mbox{$N$(\Bee)} and \mbox{$N$(\BToJpsiK)} their yields. 
The normalization mode branching fraction (including that for the decay $\decay{\jpsi}{\epem}$) is \mbox{$\mathcal{B}(\BToJpsiK)=\num{6.03\pm0.17e-5}$}, taken from Ref.\cite{PDG2019}. The \bquark-hadron fragmentation fraction ratio $f_d/f_u$ is assumed to be unity, while $f_s/f_u=\num{0.259\pm0.015}$~\cite{fsfd} is used for the Run 1 data and is scaled by $\num{1.068\pm0.016}$ for the Run 2 data, according to Ref.~\cite{LHCb-PAPER-2019-020}, to account for center-of-mass energy differences. A measurement of $f_s/f_u$ from \mbox{Run 2} yields a consistent, but less precise, result~\cite{LHCb-PAPER-2018-050}.

The yield of the normalization mode is determined using an unbinned maximum-likelihood fit to the \Kp\epem invariant mass separately for each year of data taking and bremsstrahlung category.
The fit model comprises a Gaussian function with power-law tails \cite{Skwarnicki:1986xj} for the signal component, where the tail parameters are fixed from simulation, and an exponential function to describe combinatorial background.
Summed over the bremsstrahlung categories, the yield of the normalization mode is $\num{20480\pm140}$ in the \mbox{Run 1} data and $\num{33080\pm180}$ in the Run 2 data.

The selection efficiencies $\varepsilon(\Bee)$ and $\varepsilon(\BToJpsiK)$ are determined separately for each year of data taking and bremsstrahlung category using simulated decays that are weighted to better represent the data.
Calibration data are used to evaluate particle-identification efficiencies \cite{LHCb-PUB-2016-021}. 
Trigger efficiencies are also estimated from data, using the technique described in Ref.~\cite{LHCb-PUB-2014-039}.
For simulated \Bsee decays, the mean \Bs lifetime~\cite{PDG2014} is assumed.
The selection efficiency is assumed to be the same for both \Bdee and \Bsee decays, which is consistent with results from simulation.
The normalization factors, $\alpha$, are combined across the data-taking periods and given in Table~\ref{tab:normalisation}, split by bremsstrahlung category (for the selection efficiency ratio between normalization and signal mode, see the Supplemental Material \cite{ref:supplemental}).

\begin{table}
\centering
\caption{Normalization factors $\alpha$ for \Bee. The bremsstrahlung category denotes whether zero, one or both electrons are corrected for bremsstrahlung losses. The uncertainties include statistical uncertainties and uncertainties due to limited size of the simulated samples.}
\label{tab:normalisation}
\begin{tabular}{lcc}
\toprule
Bremsstrahlung category & 2011--2012 $[10^{-5}]$& 2015--2016 $[10^{-5}]$\\
						\midrule
No correction			& $2.85 \pm 0.24$ & $1.84 \pm 0.08$     \\
One electron corrected	& $1.13 \pm 0.08$ & $0.70 \pm 0.03$	\\
Both electrons corrected	& $1.73 \pm 0.20$ & $1.04 \pm 0.06$	    \\
\bottomrule
\end{tabular}
\end{table}

In addition to the combinatorial background, backgrounds due to misidentification and partial reconstruction are present in the data. These backgrounds differ significantly between the categories of bremsstrahlung correction. Their invariant-mass shapes and relative contributions are evaluated using simulation.
In the lower mass region, partially reconstructed backgrounds of the types $\decay{\B}{X\epem}$ and $\decay{\Bu}{\Dzb(\rightarrow Y^+\en\neueb)\ep\neue}$ dominate, where $X$ and $Y$ represent hadronic systems.
The main source of background in the \B-mass region, however, stems from misidentified particles in the decays $\decay{\Bd}{\pim\ep\neue}$ and $\decay{\B}{\hadron^+\hadron^{\prime-}}$, where $h$ and $h'$ are hadrons. The latter has a peaking structure in the \B-mass region.
Backgrounds involving misidentified particles contribute mostly to categories in which at most one of the electrons has a bremsstrahlung correction applied.
The contribution from combinatorial background is evaluated from same-sign lepton pairs in data and found to be small.
The yields of the backgrounds are Gaussian constrained to their expected values, estimated from simulation using their known branching fractions~\cite{PDG2019}.

The shape of the invariant mass of the \Bsee and \Bdee components is modeled using a Gaussian function with power-law tails, where the parameters are obtained from simulation and differ between each bremsstrahlung category and year of data taking.
The peak values and the widths of the functions are corrected for data-simulation differences by a factor determined from the normalization mode.
The parameters of the \Bsee and \Bdee line shapes are fixed to the same values with the exception of the peak value, which is shifted according to the known \Bs--\Bz mass difference~\cite{PDG2019}.
Due to the limited mass resolution, arising from imperfect bremsstrahlung recovery, the line shapes from \Bsee and \Bdee are highly overlapping.
Therefore the branching fraction of \Bsee is obtained by performing a simultaneous fit to the dielectron invariant-mass distribution of all six data sets while neglecting the contribution from \Bdee, and vice versa. In these fits, the only shared parameters between categories are the branching fractions $\mathcal{B}(\Bee)$ and $\mathcal{B}(\BToJpsiK)$, and the ratio of the fragmentation fractions $f_s/f_u$.

Systematic uncertainties are estimated separately for each data set.
Dominant sources of systematic uncertainties in the normalization arise from the uncertainty on the fragmentation fraction ratio, the technique used to evaluate the trigger efficiencies, and the determination of particle-identification efficiencies;
the systematic uncertainties from these sources extend to $\SI{5.8}{\percent}$, $\SI{5.3}{\percent}$, and $\SI{5.3}{\percent}$ on the branching fractions, respectively.
The uncertainty on $\mathcal{B}(\BToJpsiK)$ of $\SI{2.8}{\percent}$~\cite{PDG2019} is taken into account.
A difference of up to $\SI{4.1}{\percent}$ is found between the efficiency of the BDT selection on simulated \BToJpsiK decays and \BToJpsiK decays in data, which is assigned as a systematic uncertainty. The fraction of candidates in each bremsstrahlung-correction category of the signal modes is taken from simulation. The difference between simulation and data is investigated using \BToJpsiK decays and its effect on the normalization, up to $\SI{4.0}{\percent}$, is taken as a systematic uncertainty.
Systematic uncertainties on the invariant-mass resolution corrections are determined by repeating the correction procedure with pseudoexperiments obtained with the bootstrapping method \cite{bootstrap}, yielding up to $\SI{1.1}{\percent}$.
A difference between the total selection efficiencies in the \Bsee and \Bdee channels of up to $\SI{2.5}{\percent}$ is assigned as a systematic uncertainty on the \Bdee normalization factor.
Due to the presence of an additional kaon in the final state of the normalization mode, the track-reconstruction efficiency is different between the signal and normalization modes. An uncertainty of $\SI{1.1}{\percent}$ is assigned to the branching fraction as a systematic uncertainty on the kaon reconstruction efficiency arising from the limited knowledge of the interactions in the detector material~\cite{LHCb-DP-2013-002}.
Finally, an uncertainty of $\SI{1.0}{\percent}$ is assigned to account for small differences in detector occupancy between the signal and normalization mode arising from the trigger selection.
The dominant sources of systematic uncertainties on the background composition are due to the imprecise knowledge of the branching fractions of the background components.
The largest uncertainty of this type on the expected background yield in the \B-mass region is $\SI{14}{\percent}$, determined from refitting the mass sidebands while varying the background components according to their uncertainties.
Taking all correlations into account, overall single event sensitivities of $[4.71\pm0.12\text{(stat.)}\pm0.33\text{(syst.)}]\times10^{-10}$ for \Bsee and $[1.271\pm0.034\text{(stat.)}\pm0.063\text{(syst.)}]\times10^{-10}$ for \Bdee are obtained.

The dielectron invariant-mass spectrum, summed over bremsstrahlung categories, is shown in Fig.~\ref{fig:massfits}, with the result of the \Bsee fit. The individual categories are shown in the Supplemental Material~\cite{ref:supplemental}, as well as the distributions with the result of the \Bdee fit.
The measured branching fractions are \mbox{$\mathcal{B}(\Bsee) = \num{2.4 \pm 4.4e-9}$} and \mbox{$\mathcal{B}(\Bdee) = \num{0.30 \pm 1.29e-9}$}, where the uncertainties include both statistical and systematic components. The results are in agreement with the background-only hypothesis.

\begin{figure}
\centering
\includegraphics[width=0.49\textwidth]{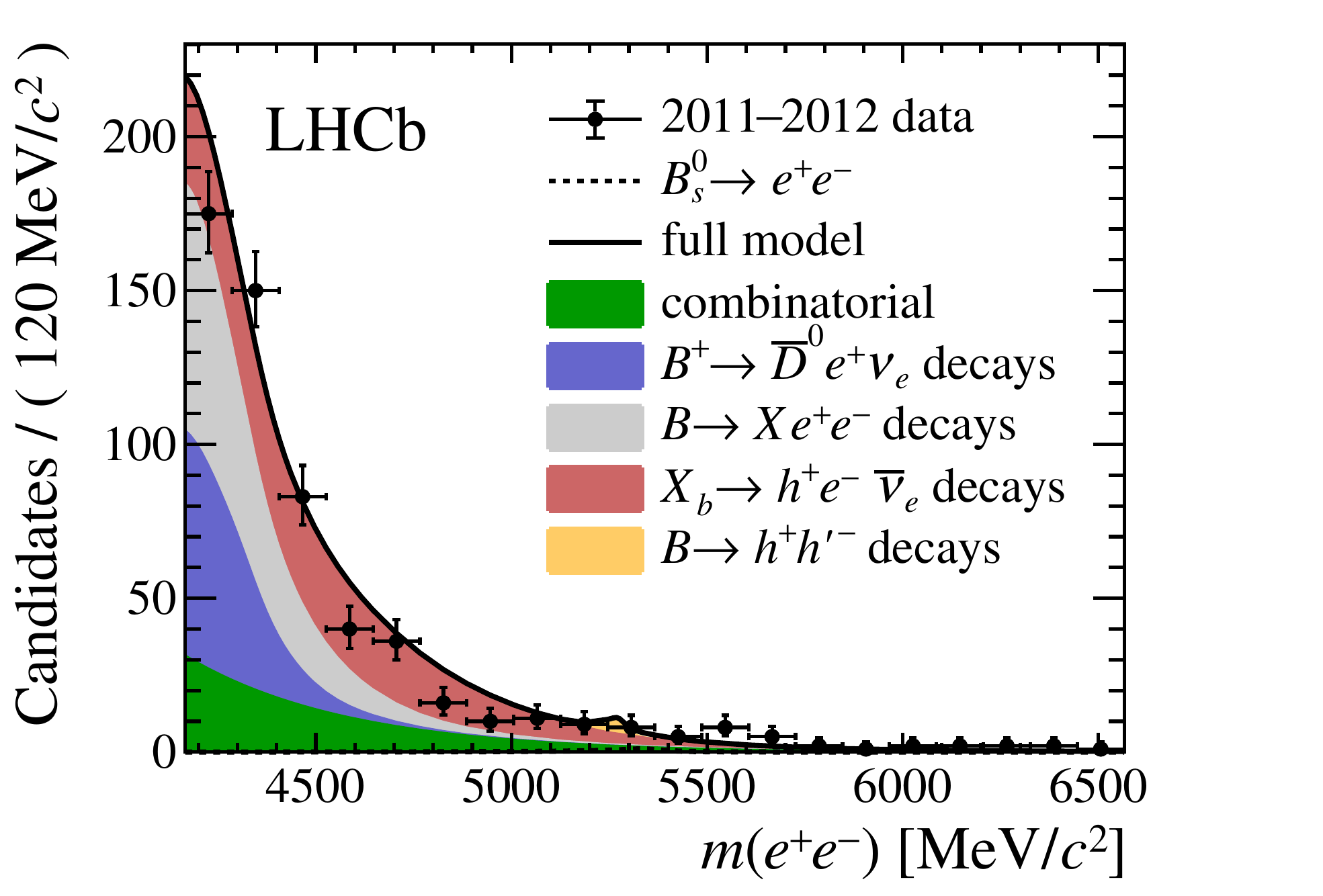}
\includegraphics[width=0.49\textwidth]{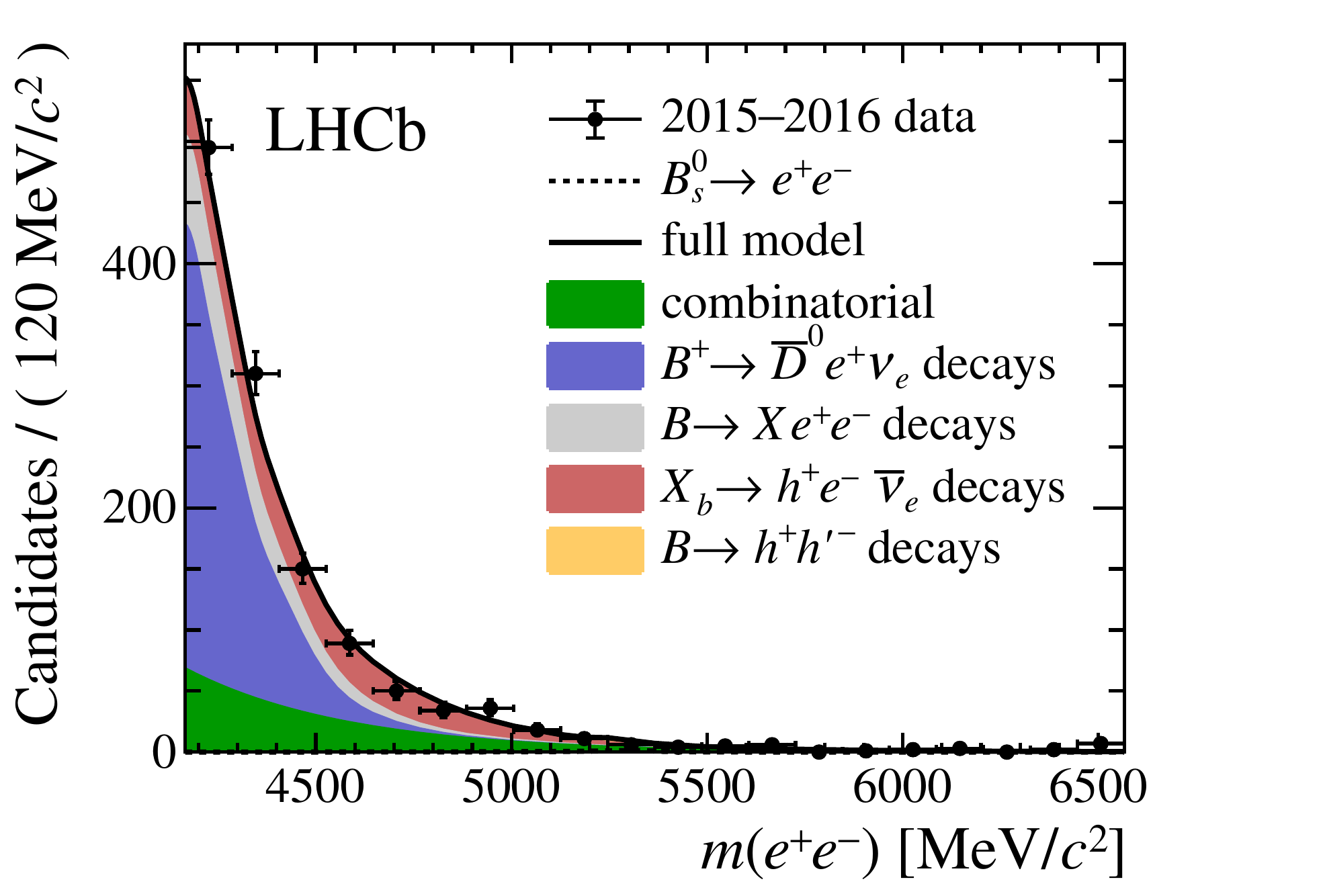}
\caption{Simultaneous fit to the dielectron invariant-mass distribution, with $\mathcal{B}(\Bdee)$ fixed to zero. The sum of bremsstrahlung categories is shown for (left) Run 1 and (right) Run 2. The relative proportions of background contributions change between Run 1 and Run 2 due to different performances of the particle identification algorithms and BDT selections.}
\label{fig:massfits}
\end{figure}

Upper limits on the branching fractions are set using the CL$_s$ method~\cite{CLs}, as implemented in the \textsc{GammaCombo} framework\cite{LHCb-PAPER-2016-032,GammaCombo} with a one-sided profile likelihood ratio \cite{Cowan:2010js} as test statistic.
The likelihoods are computed from fits to the invariant-mass distributions.
In the fits, the normalization factor, normalization mode branching fraction, fragmentation fraction ratio, and background yields are Gaussian constrained to their expected values  within statistical and systematic uncertainties. Pseudoexperiments, in which the nuisance parameters are set to their fitted values from data, are used for the evaluation of the test statistic.

\begin{figure}
\centering
\includegraphics[width=0.49\textwidth]{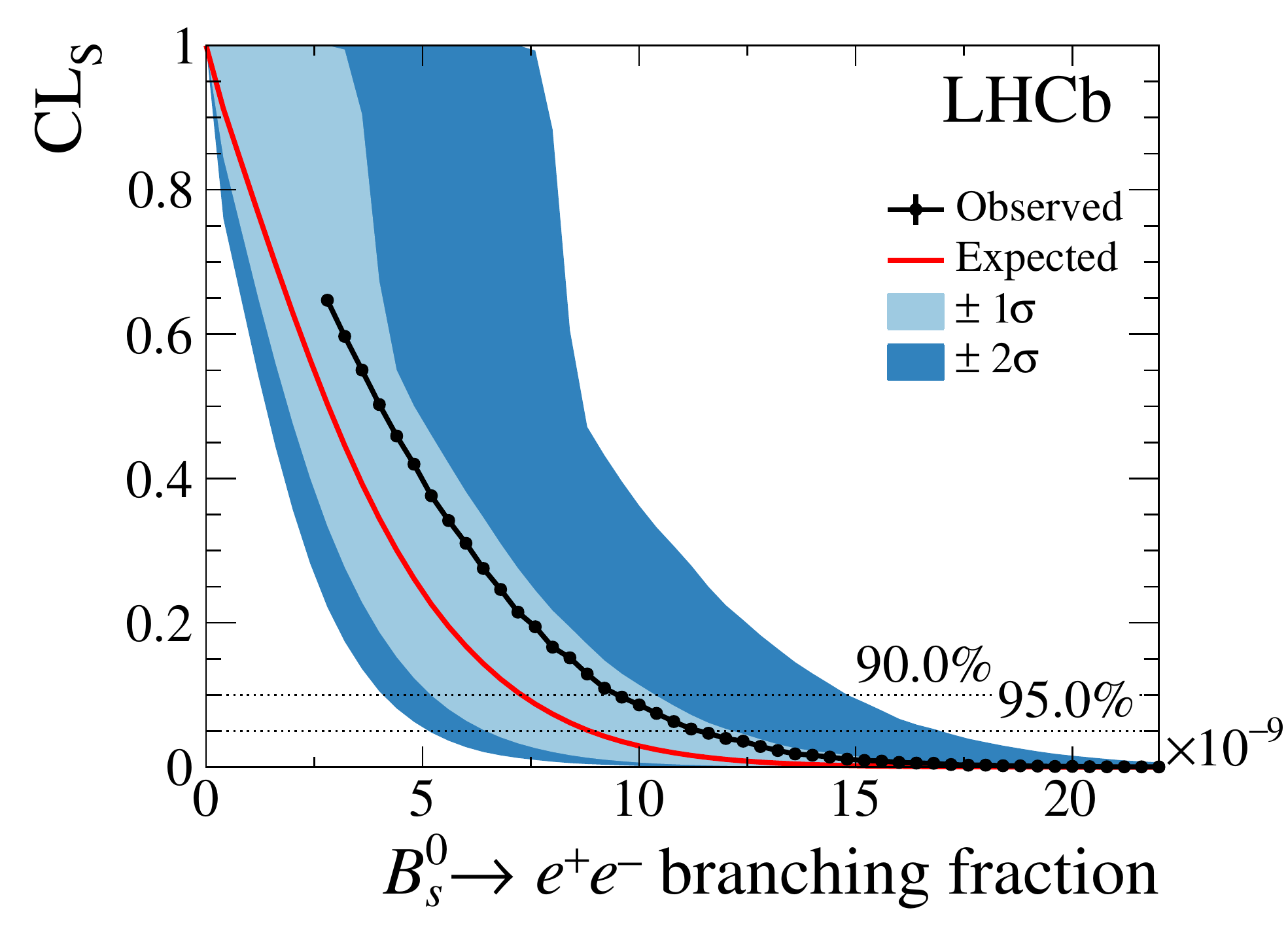}
\includegraphics[width=0.49\textwidth]{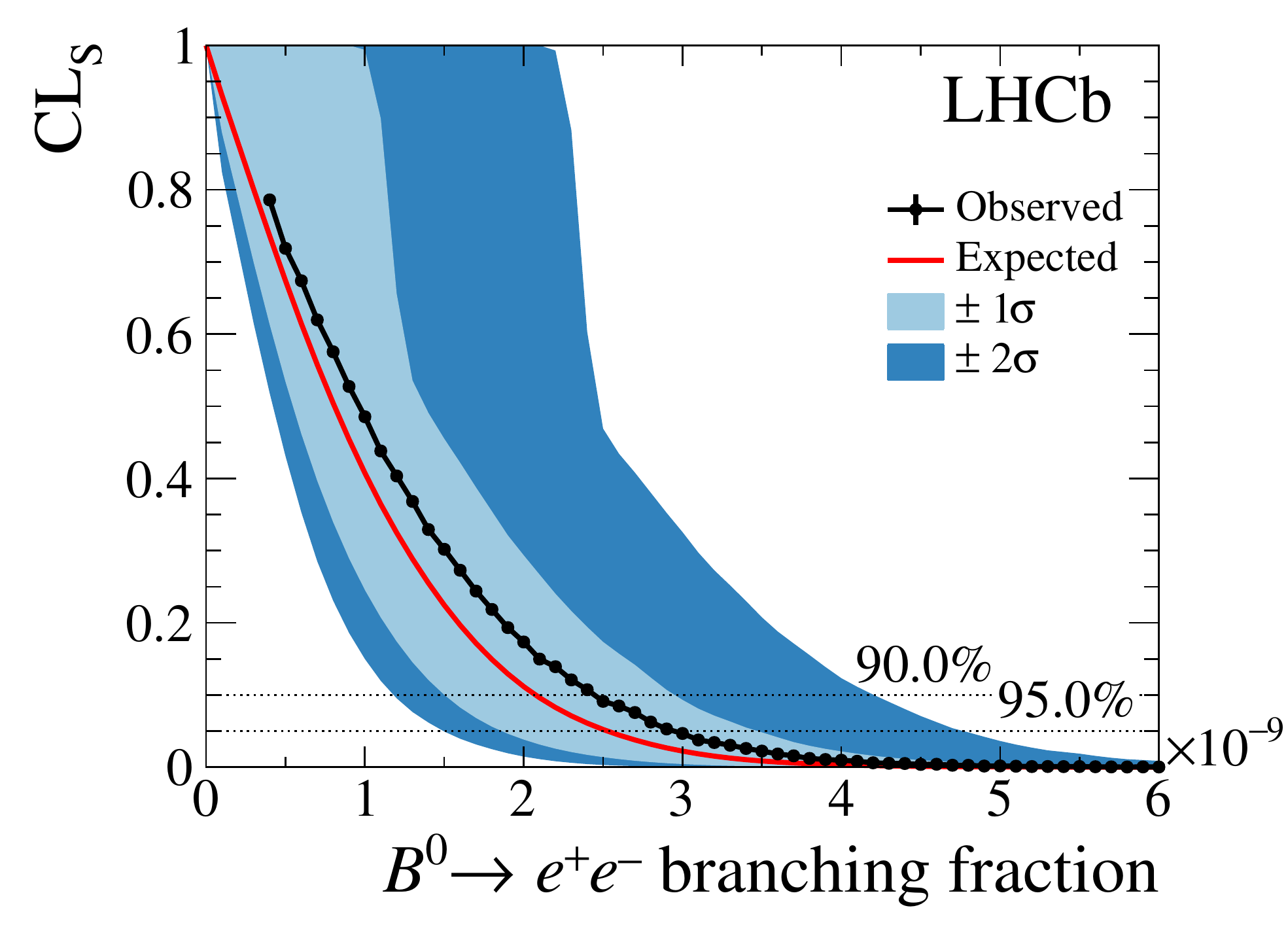}
\caption{CL$_s$ values as a function of the branching fractions of the decays (left) \Bsee and (right) \Bdee. The red solid line (black solid line with data points) corresponds to the distribution of the expected
(observed) upper limits, and the light blue (dark blue) band contains the $1\sigma$ $(2\sigma)$ uncertainties on the expected upper limits. Thresholds corresponding to $90\,\%$ and $95\,\%$ confidence level are indicated with dashed lines. The observed values are plotted for branching fractions greater than the measured branching fraction in the data; the test statistic is defined to be nonzero only in that region.}
\label{fig:clsplots}
\end{figure}

The expected and observed CL$_s$ distributions are shown in Fig.~\ref{fig:clsplots}. The upper observed limits are $\mathcal{B}(\Bsee)<9.4\,(11.2)\times10^{-9}$ and $\mathcal{B}(\Bdee)<2.5\,(3.0)\times10^{-9}$ at 
$90\,(95)\,\%$ confidence level. These are consistent with the expected upper limits of \mbox{$\mathcal{B}(\Bsee)<7.0\,(8.6)\times10^{-9}$} and \mbox{$\mathcal{B}(\Bdee)<2.0\,(2.5)\times10^{-9}$} at 
$90\,(95)\,\%$ confidence level, obtained as the median of limits determined on background-only pseudoexperiments.

In conclusion, a search for the rare decays \Bee is performed using data from proton-proton collisions recorded with the LHCb experiment, corresponding to a total integrated luminosity of $\SI{5}{\invfb}$. No excess of events is observed over the background. The resulting limits on the branching fractions are \mbox{$\mathcal{B}(\Bsee)<9.4\,(11.2)\times10^{-9}$} and \mbox{$\mathcal{B}(\Bdee)<2.5\,(3.0)\times10^{-9}$} at 
$90\,(95)\,\%$ confidence level, when neglecting the contribution from the other decay. The mean \Bs lifetime is assumed for \Bsee decays. Assuming SM-like \CP-odd (\CP-even) \Bsee decays, an increase (decrease) of $\SI{2.4}{\percent}$ with respect to the quoted limit is found. The results improve the limits on these branching fractions~\cite{Aaltonen:2009vr} by more than one order of magnitude and constrain contributions beyond the SM, for example from scalar and pseudoscalar currents~\cite{Fleischer:2017ltw}.

\section*{Acknowledgements}
\noindent We express our gratitude to our colleagues in the CERN
accelerator departments for the excellent performance of the LHC. We
thank the technical and administrative staff at the LHCb
institutes.
We acknowledge support from CERN and from the national agencies:
CAPES, CNPq, FAPERJ and FINEP (Brazil); 
MOST and NSFC (China); 
CNRS/IN2P3 (France); 
BMBF, DFG and MPG (Germany); 
INFN (Italy); 
NWO (Netherlands); 
MNiSW and NCN (Poland); 
MEN/IFA (Romania); 
MSHE (Russia); 
MinECo (Spain); 
SNSF and SER (Switzerland); 
NASU (Ukraine); 
STFC (United Kingdom); 
DOE NP and NSF (USA).
We acknowledge the computing resources that are provided by CERN, IN2P3
(France), KIT and DESY (Germany), INFN (Italy), SURF (Netherlands),
PIC (Spain), GridPP (United Kingdom), RRCKI and Yandex
LLC (Russia), CSCS (Switzerland), IFIN-HH (Romania), CBPF (Brazil),
PL-GRID (Poland) and OSC (USA).
We are indebted to the communities behind the multiple open-source
software packages on which we depend.
Individual groups or members have received support from
AvH Foundation (Germany);
EPLANET, Marie Sk\l{}odowska-Curie Actions and ERC (European Union);
ANR, Labex P2IO and OCEVU, and R\'{e}gion Auvergne-Rh\^{o}ne-Alpes (France);
Key Research Program of Frontier Sciences of CAS, CAS PIFI, and the Thousand Talents Program (China);
RFBR, RSF and Yandex LLC (Russia);
GVA, XuntaGal and GENCAT (Spain);
the Royal Society
and the Leverhulme Trust (United Kingdom).

\clearpage
\newpage
\section*{Supplemental Material for LHCb-PAPER-2020-001}
\label{sec:Supplemental}
\begin{figure}[htbp]
\centering
\includegraphics[width=0.328\textwidth]{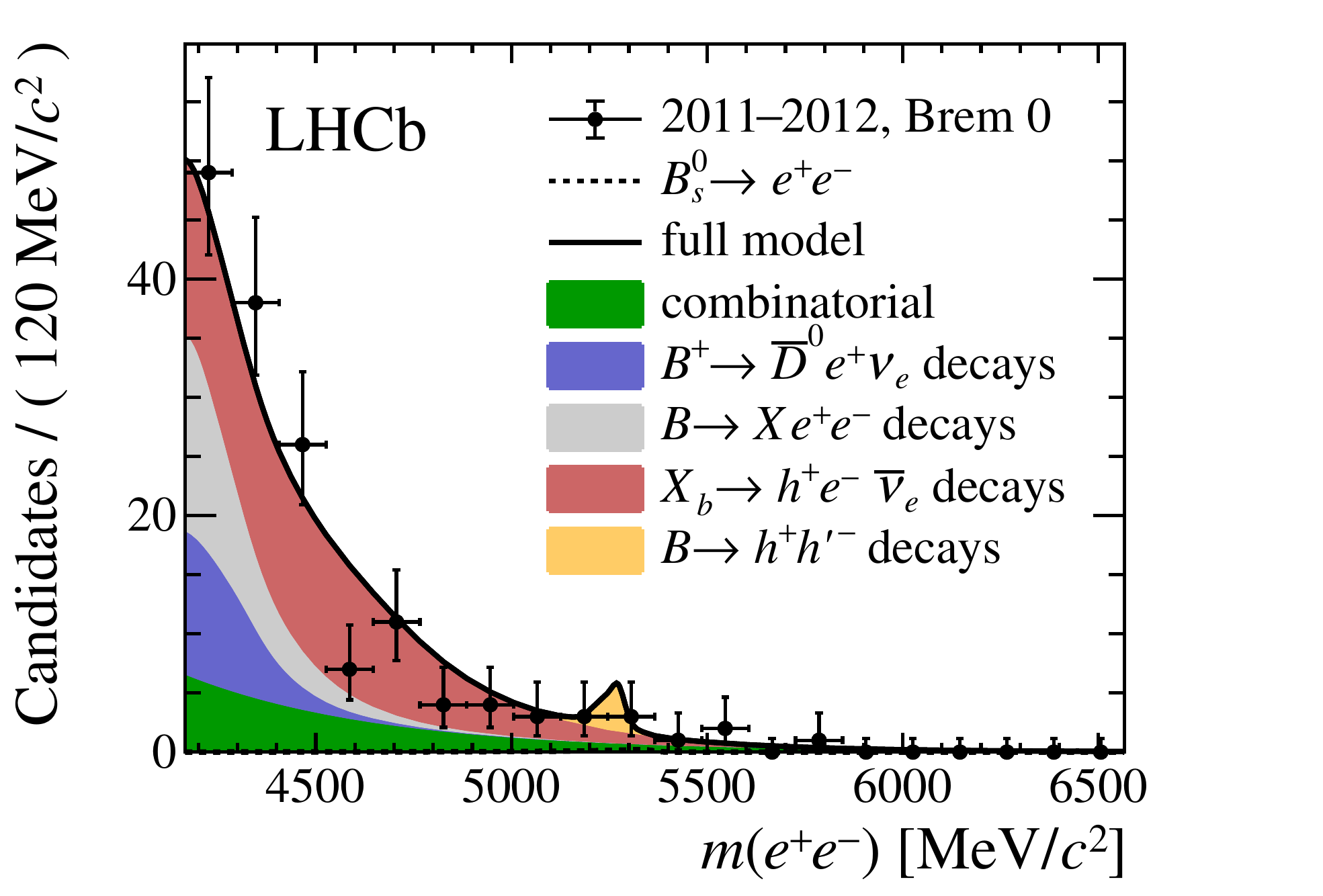}
\includegraphics[width=0.328\textwidth]{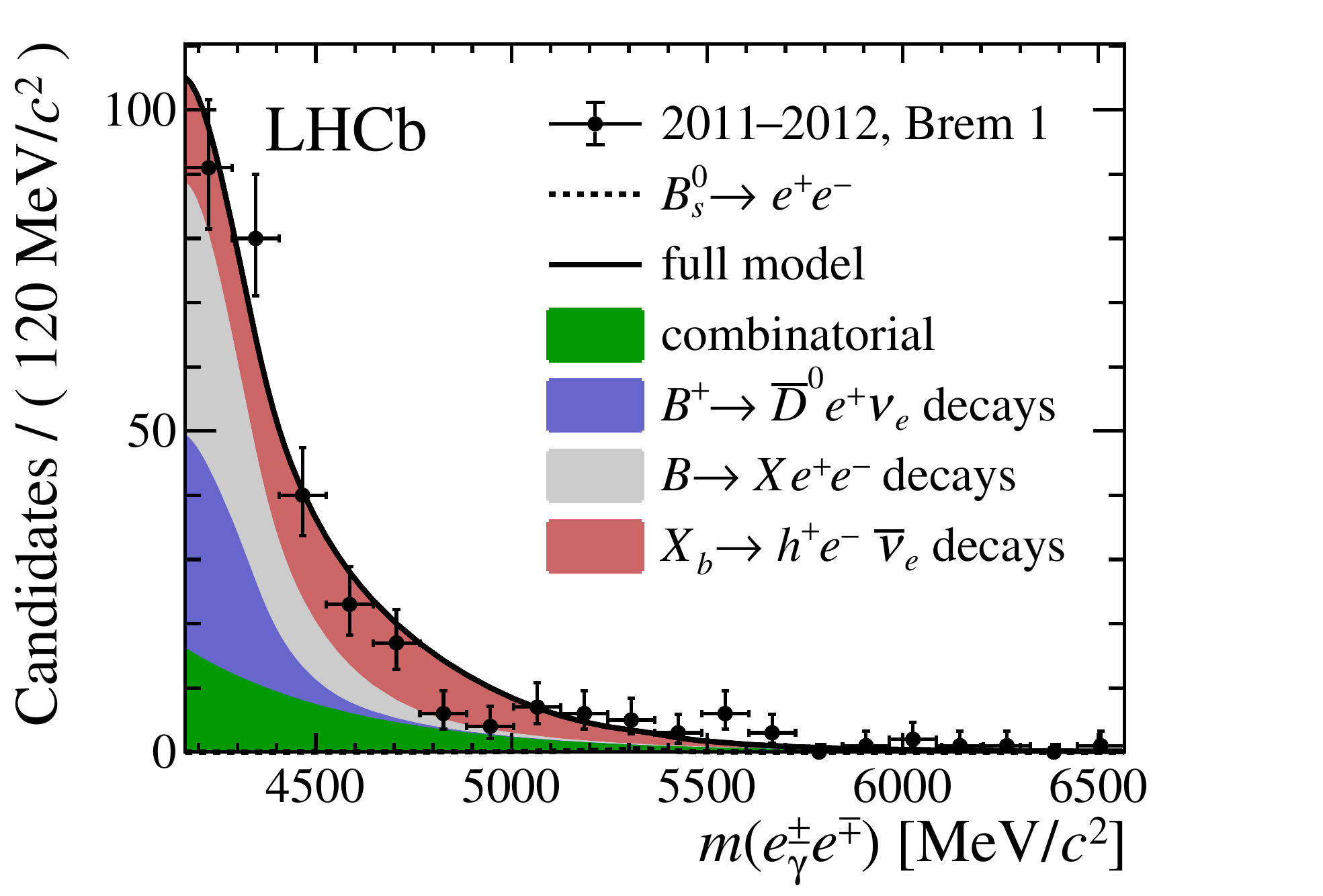}
\includegraphics[width=0.328\textwidth]{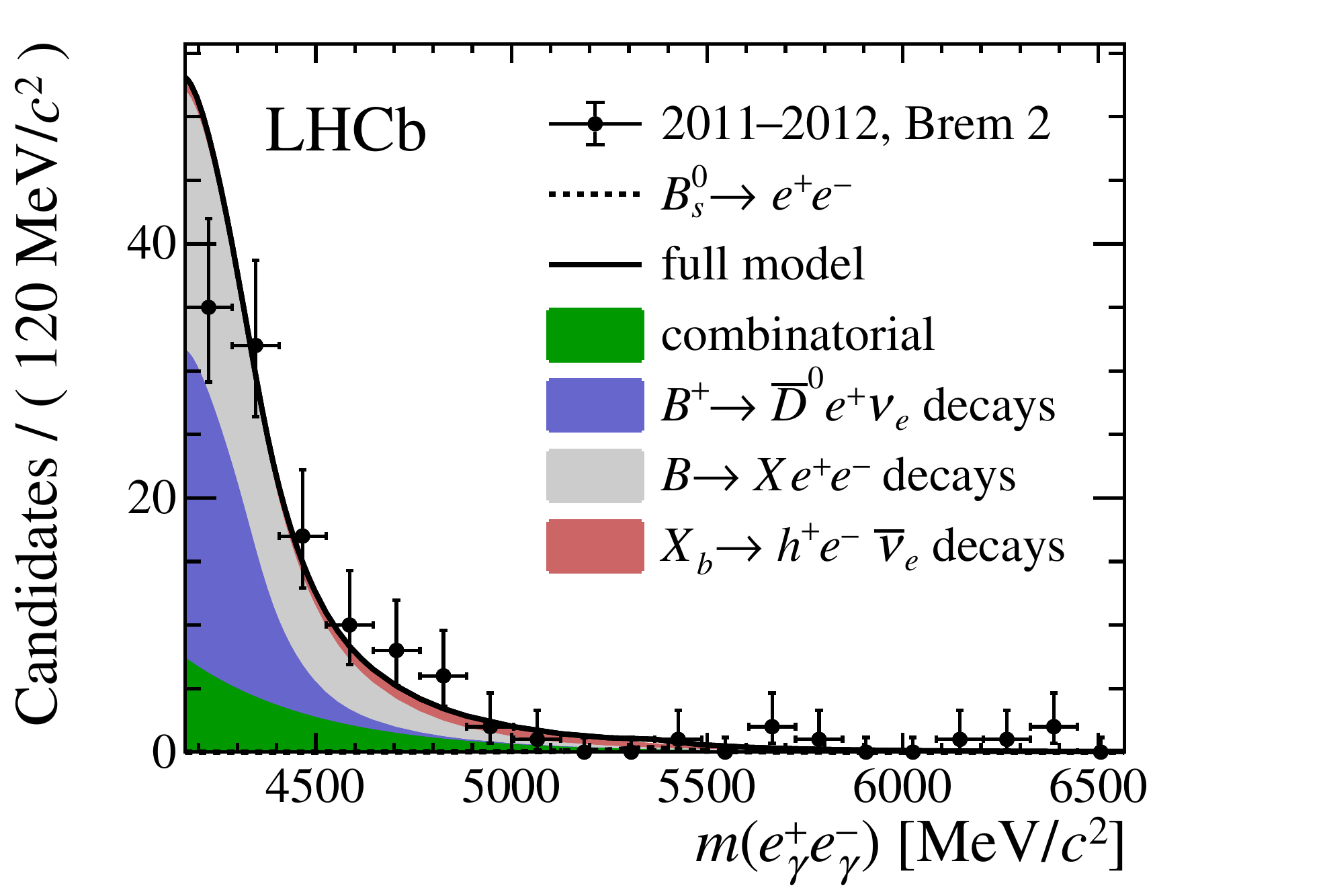}\\
\includegraphics[width=0.328\textwidth]{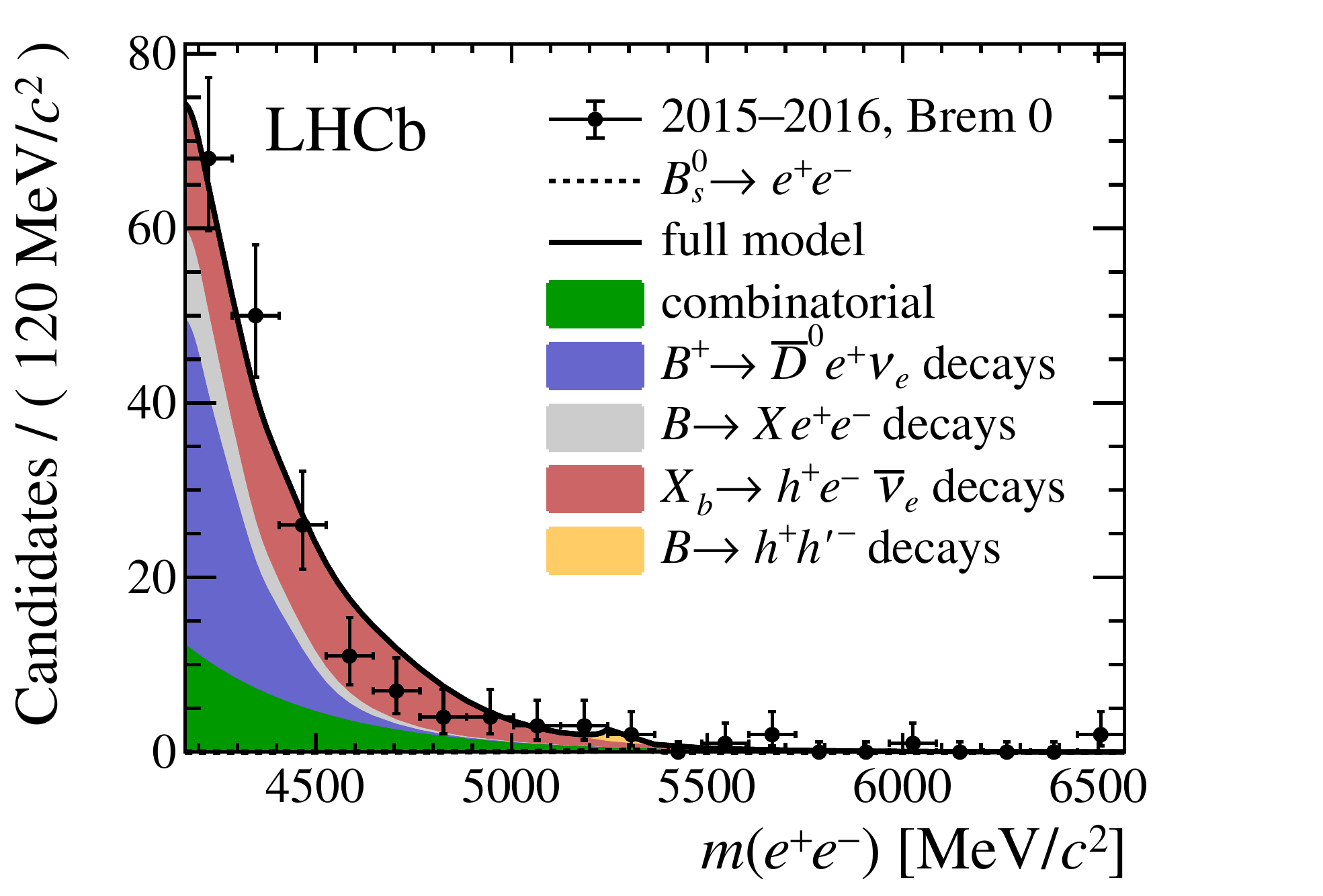}
\includegraphics[width=0.328\textwidth]{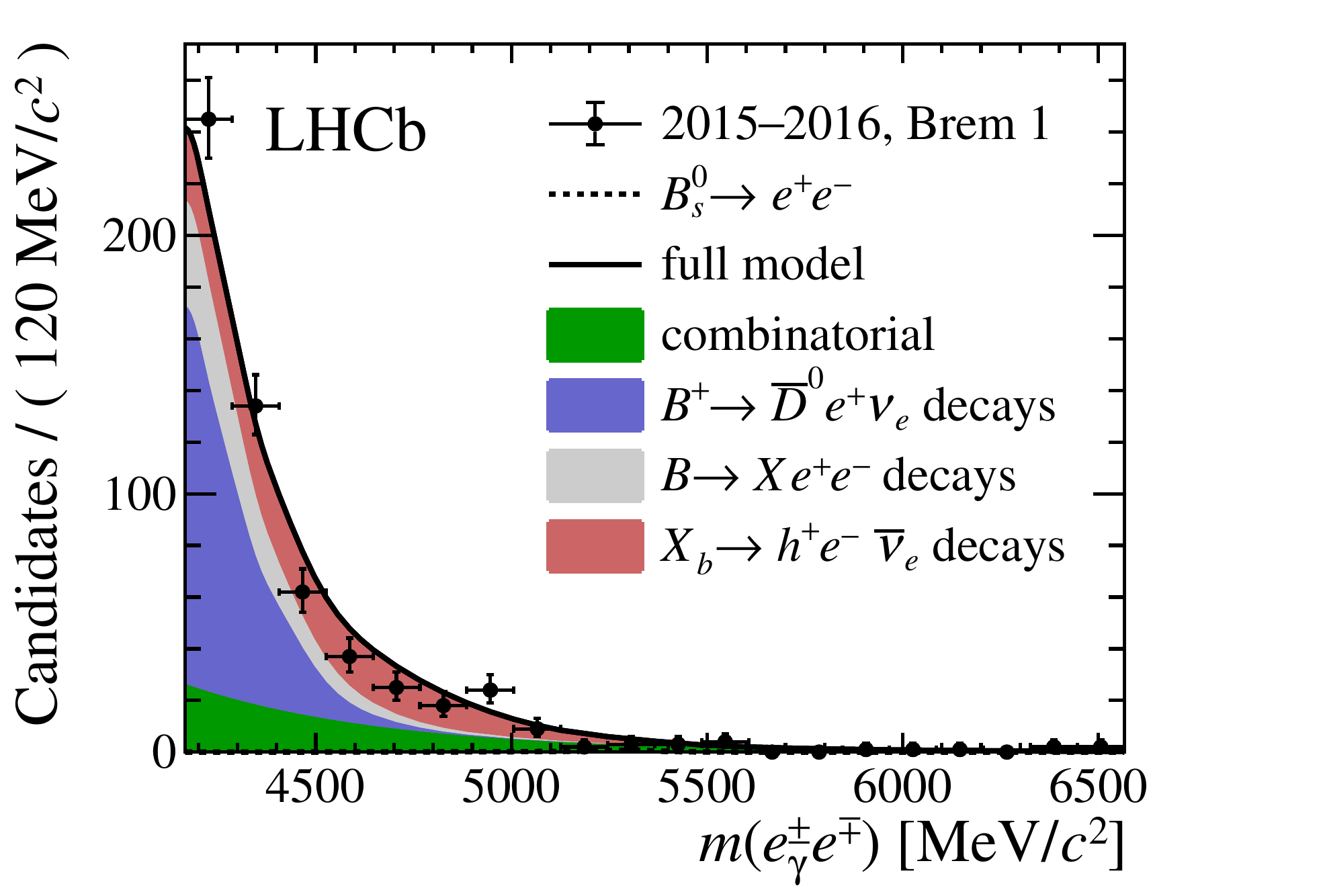}
\includegraphics[width=0.328\textwidth]{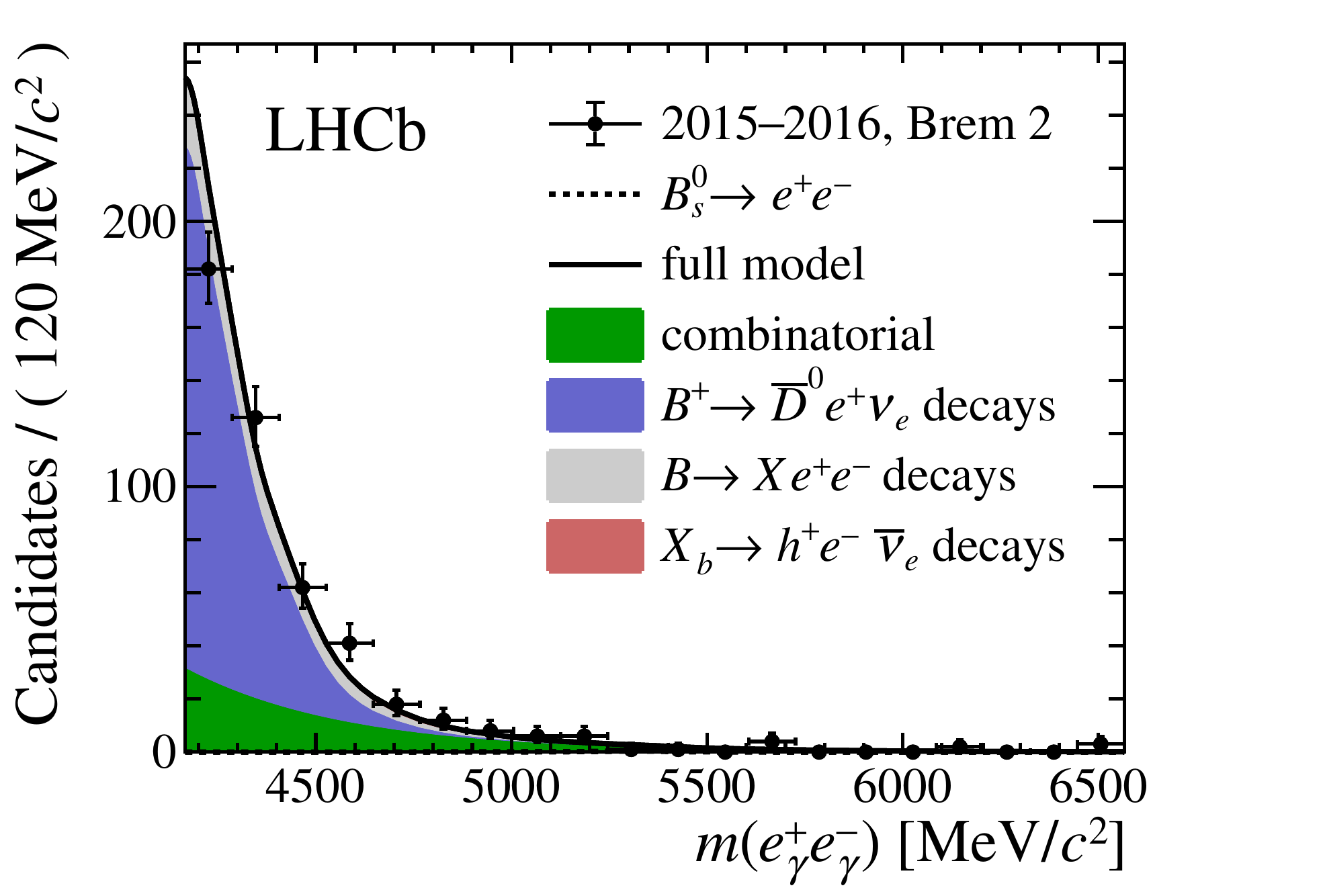}
\caption{Simultaneous fit to the dielectron invariant-mass distribution in all categories, with $\mathcal{B}(\Bdee)$ fixed to zero. The top figures show the three bremsstrahlung categories in the Run 1 data set and the bottom figures show the Run 2 data set. From left to right, the data sets correspond to the bremsstrahlung correction category with no correction, correcting one electron and correcting both electrons. The relative proportions of background contributions change between Run 1 and Run 2 due to different performances of the particle-identification algorithms and BDT selections. Their relative fractions between bremsstrahlung categories follow the expectation from simulation.}
\label{fig:massfits_Bsee}
\end{figure}

\begin{figure}[htbp]
\centering
\includegraphics[width=0.49\textwidth]{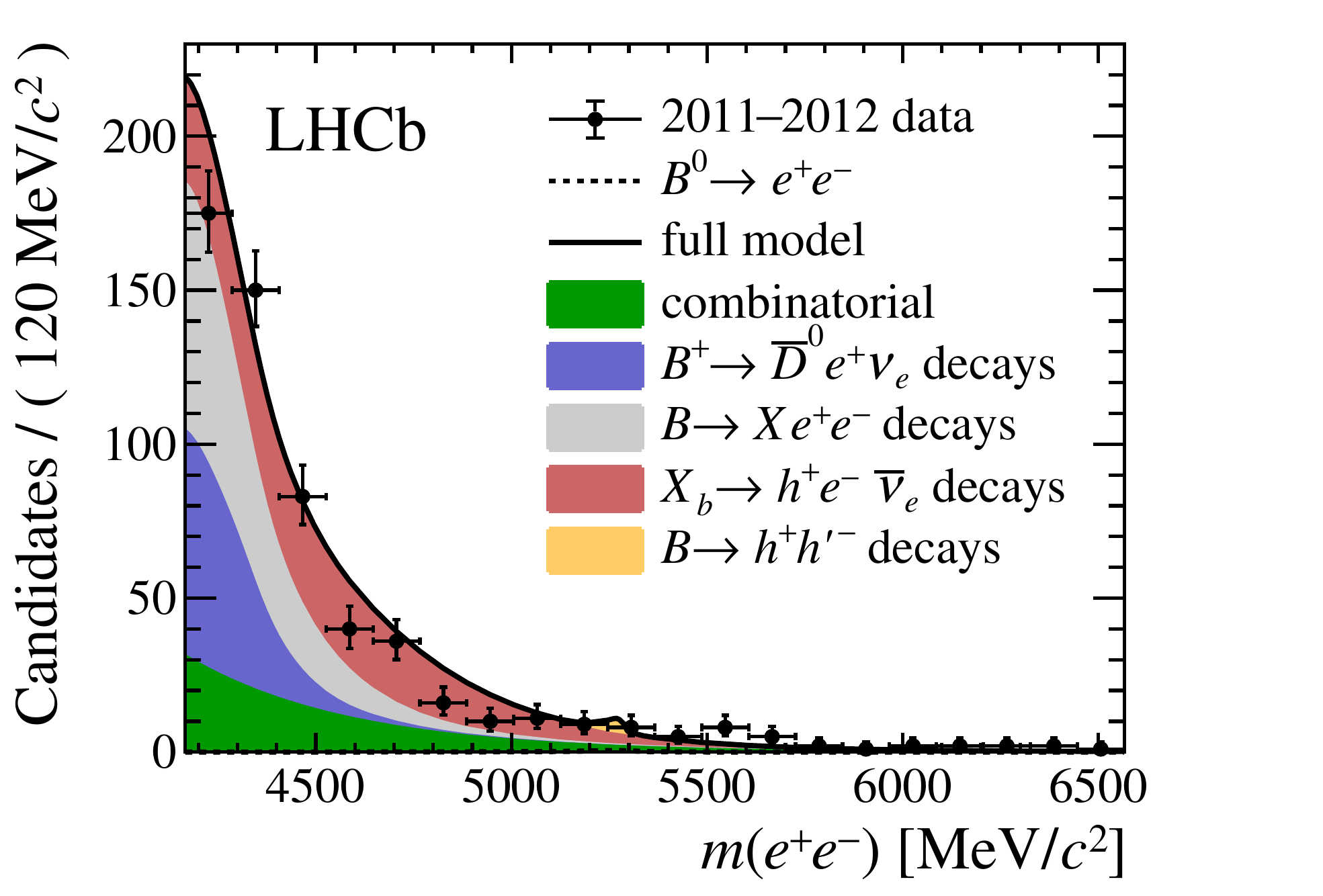}
\includegraphics[width=0.49\textwidth]{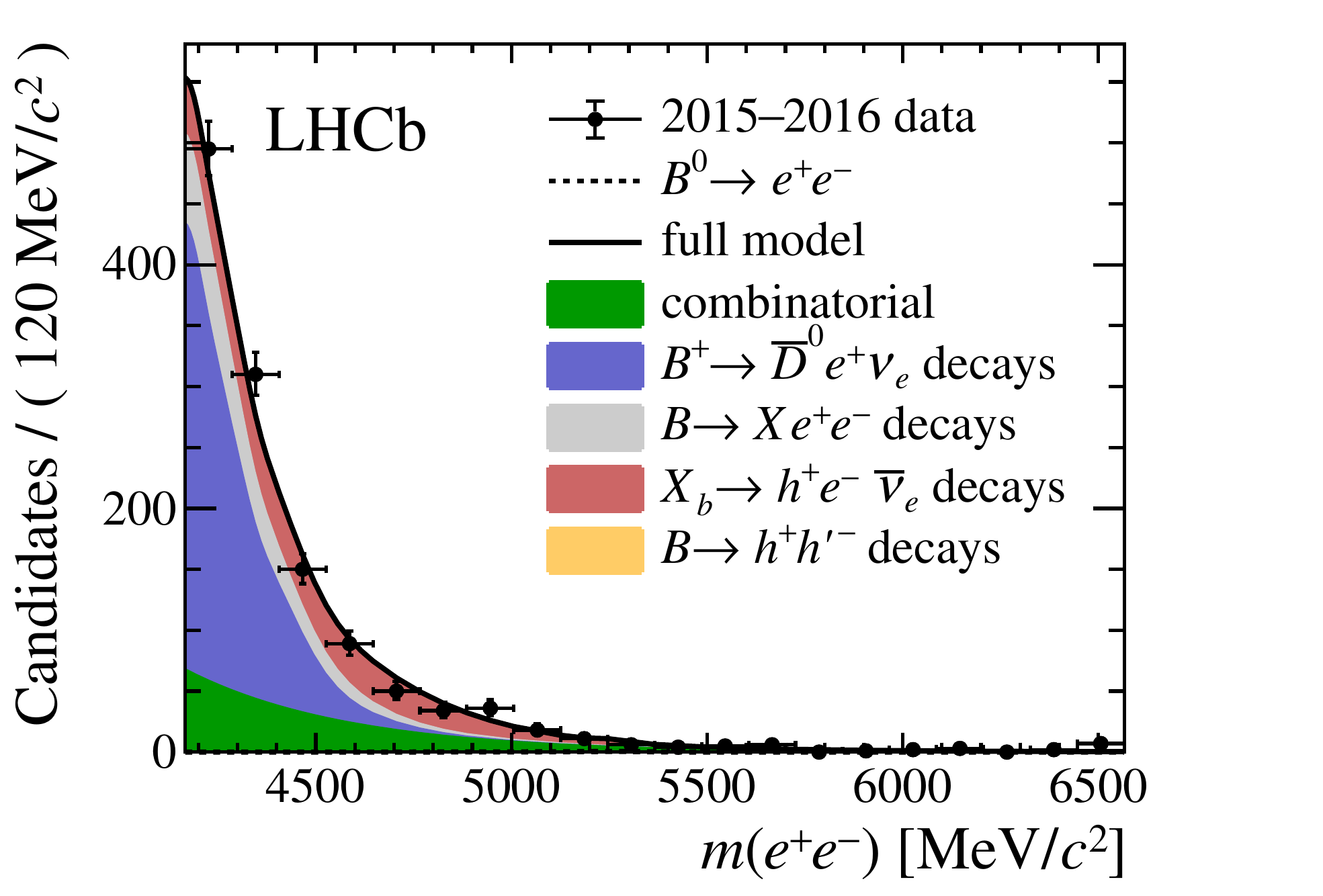}
\caption{Simultaneous fit to the dielectron invariant-mass distribution, with $\mathcal{B}(\Bsee)$ fixed to zero. The bremsstrahlung categories are summed over the (left) Run 1 and (right) Run 2 data sets. The relative proportions of background contributions change between Run 1 and Run 2 due to different performances of the particle-identification algorithms and BDT selections.}
\label{fig:massfits_Bdee_sum}
\end{figure}

\begin{figure}[htbp]
\centering
\includegraphics[width=0.328\textwidth]{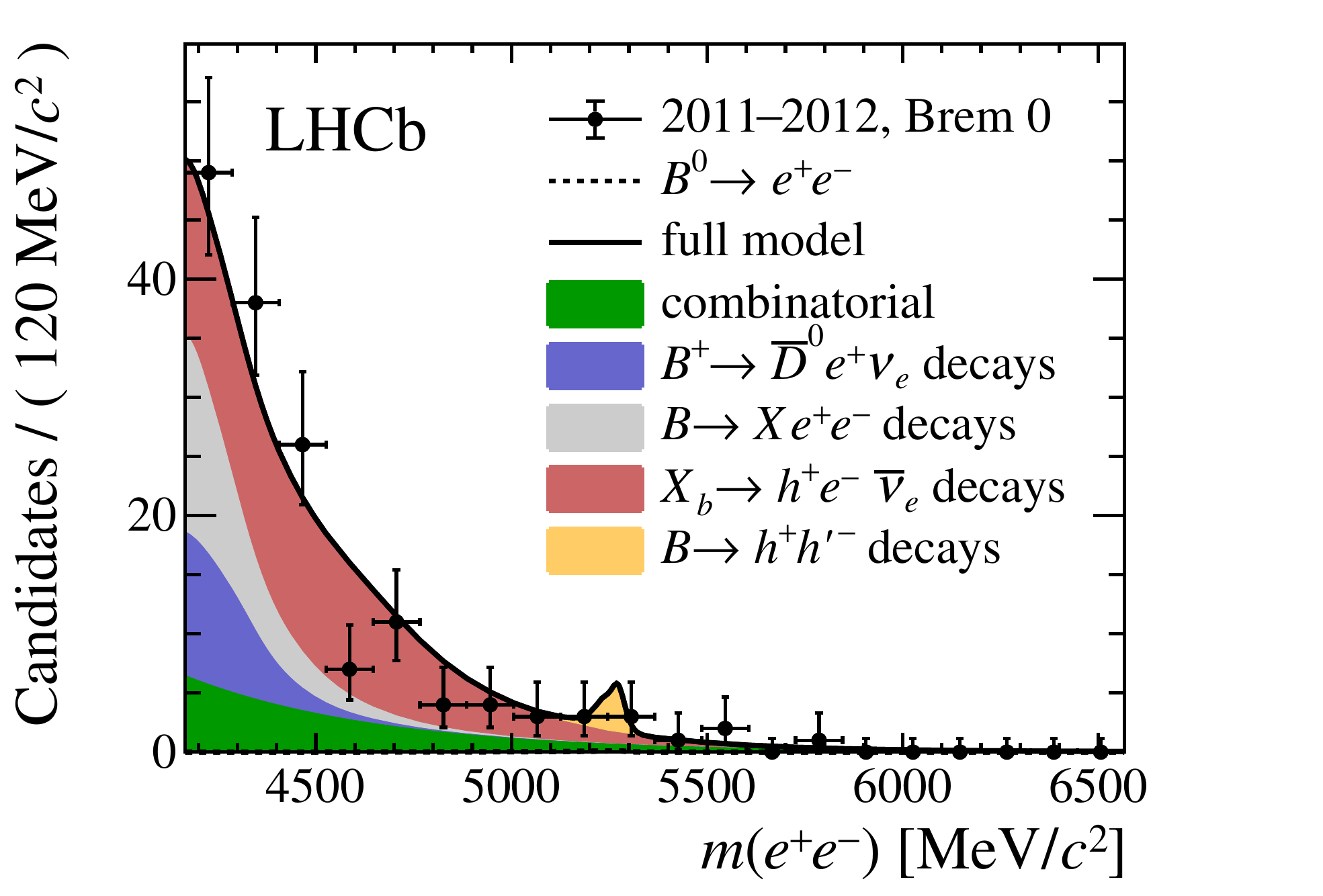}
\includegraphics[width=0.328\textwidth]{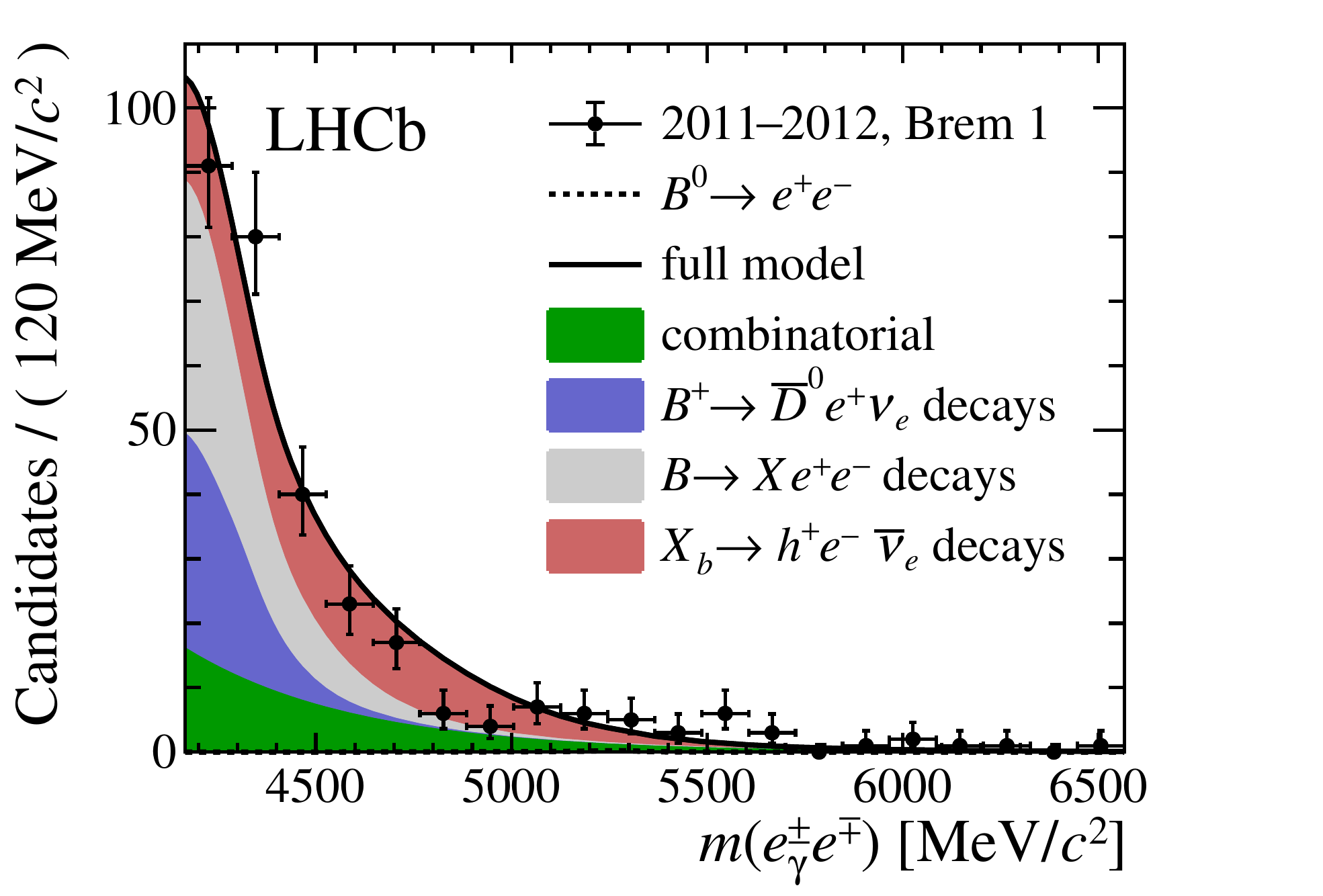}
\includegraphics[width=0.328\textwidth]{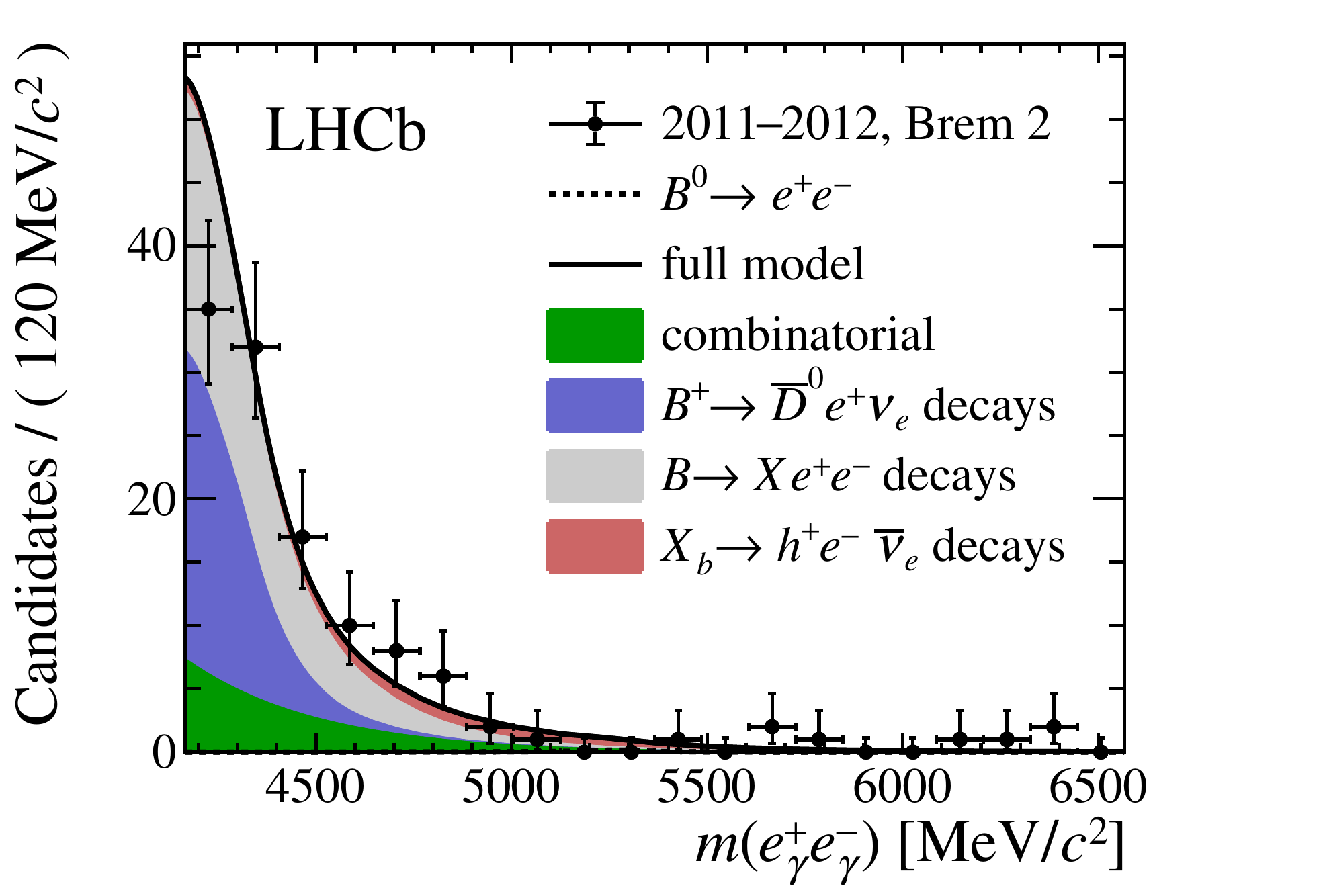}\\
\includegraphics[width=0.328\textwidth]{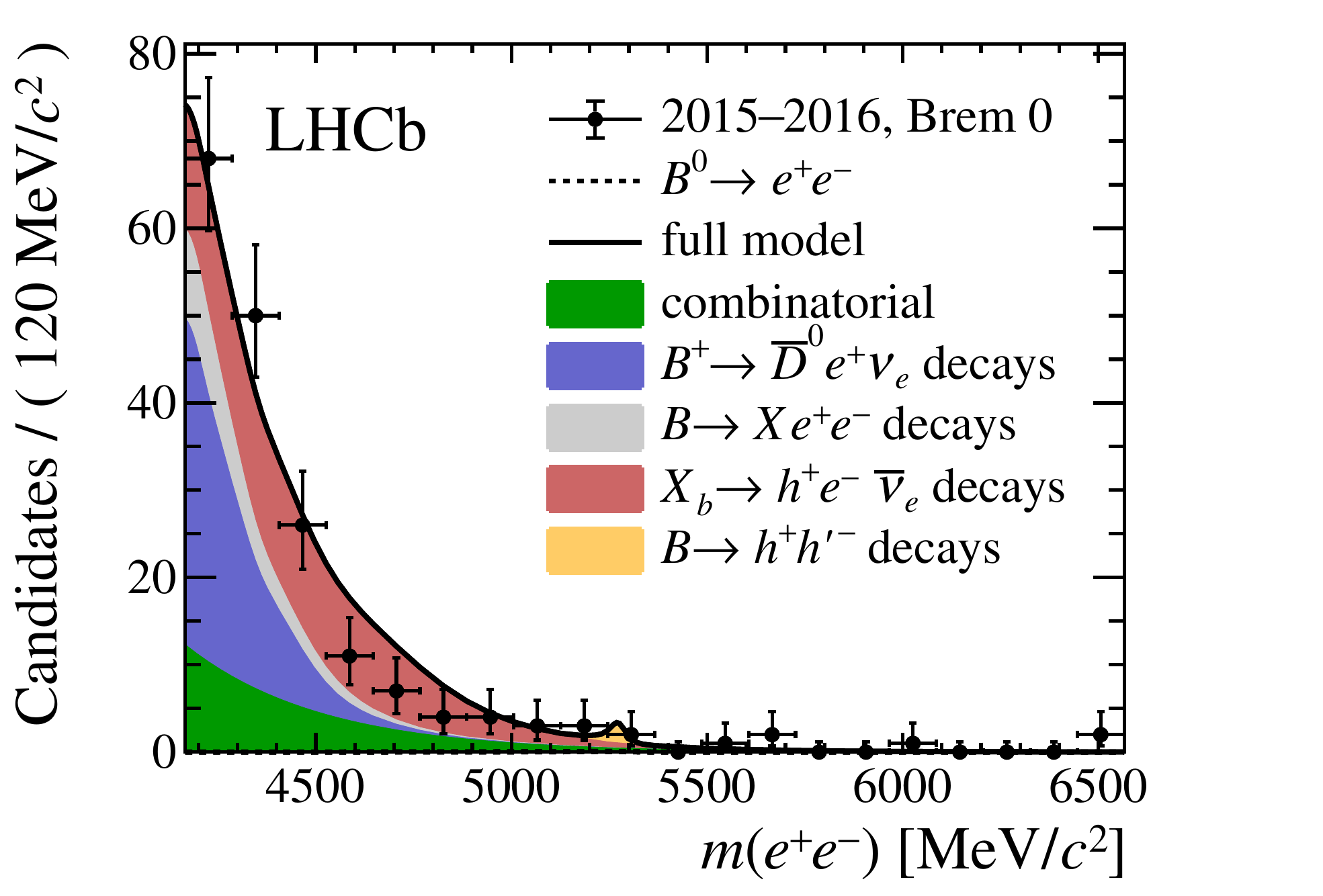}
\includegraphics[width=0.328\textwidth]{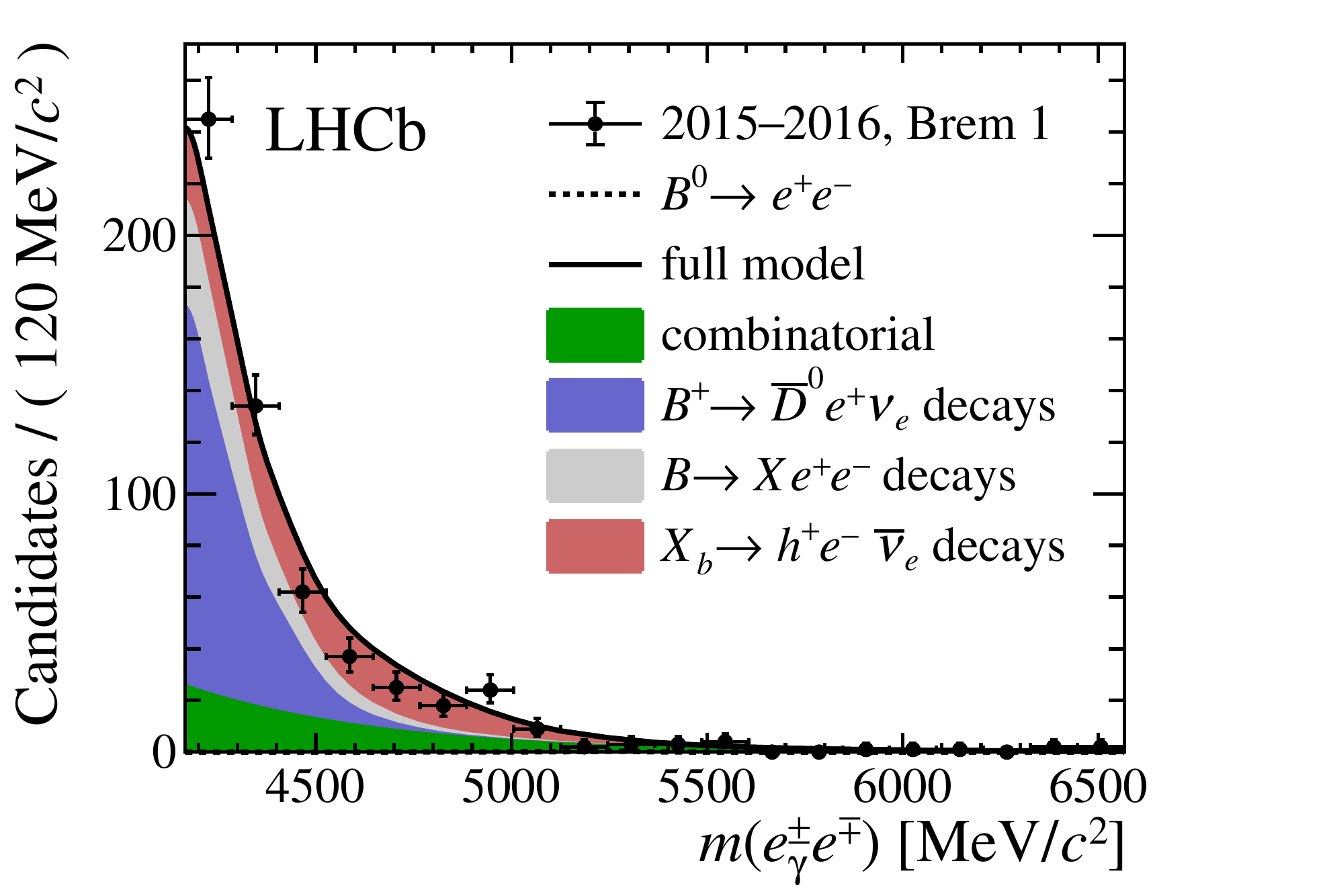}
\includegraphics[width=0.328\textwidth]{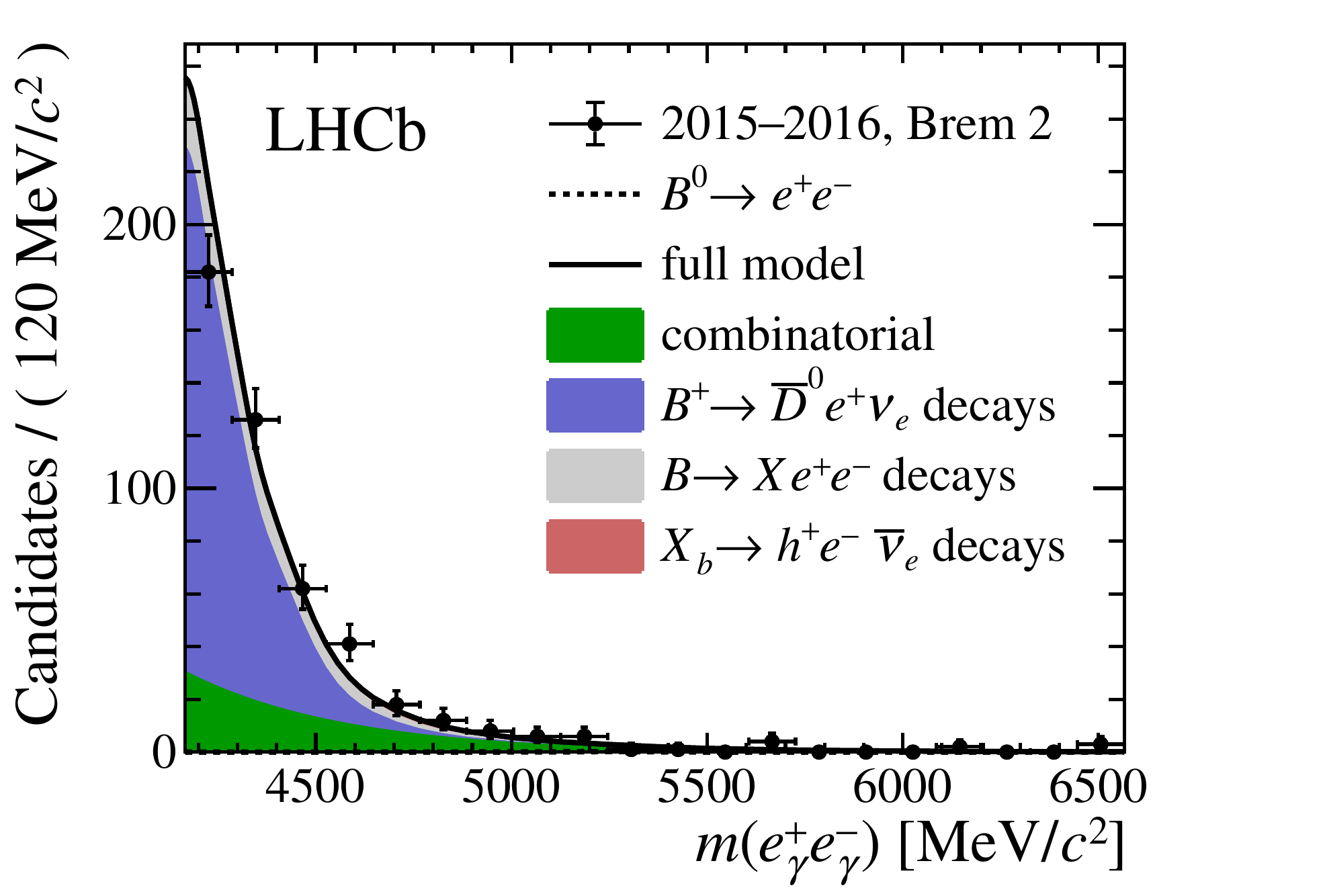}
\caption{Simultaneous fit to the dielectron invariant-mass distribution in all categories, with $\mathcal{B}(\Bsee)$ fixed to zero. The top figures show the three bremsstrahlung categories in the Run 1 data set and the bottom figures show the Run 2 data set. From left to right, the data sets correspond to the bremsstrahlung correction category with no correction, correcting one electron and correcting both electrons. The relative proportions of background contributions change between Run 1 and Run 2 due to different performances of the particle-identification algorithms and BDT selections. Their relative fractions between bremsstrahlung categories follow the expectation from simulation.}
\label{fig:massfits_Bdee}
\end{figure}

The individual categories of the simultaneous fit to the dielectron invariant-mass using the \Bsee hypothesis are presented in Fig.~\ref{fig:massfits_Bsee}.
The fit to the invariant dielectron mass including the \Bdee hypothesis instead of the \Bsee hypothesis is shown in Fig.~\ref{fig:massfits_Bdee_sum}, where the bremsstrahlung categories are summed.
The individual categories of the simultaneous fit to the dielectron invariant-mass using the \Bdee hypothesis are presented in Fig.~\ref{fig:massfits_Bdee}.
Table~\ref{tab:normalisation_inputs} lists the inputs to the normalization factors: the ratio of normalization and signal efficiencies and the normalization yield. The efficiency of the normalization mode differs from the signal and causes the efficiency ratio to decrease with bremsstrahlung category due to the slightly different reconstruction and preselection and a different impact of the BDT selection, where the differences mainly originate from the additional track in the normalization mode.

\begin{table}
\centering
\caption{Inputs for the normalization factors, the efficiency ratio \mbox{$\varepsilon(\BToJpsiK)/\varepsilon(\Bee)$} and normalization yield $N(\BToJpsiK)$. The bremsstrahlung category (Brem. cat.) denotes whether zero, one or both electrons are corrected for bremsstrahlung losses. The uncertainties on the efficiency ratios include statistical uncertainties from the calibration and uncertainties due to limited size of the simulated samples.}
\label{tab:normalisation_inputs}
\begin{tabular}{lcccc}
\toprule
Brem. cat. & \multicolumn{2}{c}{2011--2012}& \multicolumn{2}{c}{2015--2016}\\
						\midrule
						& Efficiency ratio & Norm. yield $[10^3]$ & Efficiency ratio & Norm. yield $[10^3]$\\
						\midrule
Brem. 0			& $0.144 \pm 0.012$ & $5.05 \pm 0.07$   &   $0.148\pm0.118$ & $7.96\pm0.09$  \\
Brem. 1	        & $0.119 \pm 0.008$ & $10.43 \pm 0.11$	&   $0.118\pm0.005$ & $12.75\pm0.05$ \\
Brem. 2	        & $0.086 \pm 0.010$ & $4.95 \pm 0.07$	&   $0.085\pm0.005$ & $8.306\pm0.032$ \\
\bottomrule
\end{tabular}
\end{table}

\newpage
\clearpage
\newpage

\addcontentsline{toc}{section}{References}
\setboolean{inbibliography}{true}
\ifx\mcitethebibliography\mciteundefinedmacro
\PackageError{LHCb.bst}{mciteplus.sty has not been loaded}
{This bibstyle requires the use of the mciteplus package.}\fi
\providecommand{\href}[2]{#2}


\newpage
\centerline
{\large\bf LHCb collaboration}
\begin
{flushleft}
\small
R.~Aaij$^{31}$,
C.~Abell{\'a}n~Beteta$^{49}$,
T.~Ackernley$^{59}$,
B.~Adeva$^{45}$,
M.~Adinolfi$^{53}$,
H.~Afsharnia$^{9}$,
C.A.~Aidala$^{81}$,
S.~Aiola$^{25}$,
Z.~Ajaltouni$^{9}$,
S.~Akar$^{66}$
% ...

{\footnotesize \it

% ...
$ ^{9}$Universit{\'e} Clermont Auvergne, CNRS/IN2P3, LPC, Clermont-Ferrand, France\\
% ...
$ ^{25}$INFN Sezione di Milano, Milano, Italy\\
% ...
$ ^{31}$Nikhef National Institute for Subatomic Physics, Amsterdam, Netherlands\\
% ...
$ ^{45}$Instituto Galego de F{\'\i}sica de Altas Enerx{\'\i}as (IGFAE), Universidade de Santiago de Compostela, Santiago de Compostela, Spain\\
% ...
$ ^{49}$Physik-Institut, Universit{\"a}t Z{\"u}rich, Z{\"u}rich, Switzerland\\
% ...
$ ^{53}$H.H. Wills Physics Laboratory, University of Bristol, Bristol, United Kingdom\\
% ...
$ ^{59}$Oliver Lodge Laboratory, University of Liverpool, Liverpool, United Kingdom\\
% ...
$ ^{66}$Los Alamos National Laboratory (LANL), Los Alamos, United States\\
$ ^{67}$Syracuse University, Syracuse, NY, United States\\
% ...
$ ^{81}$University of Michigan, Ann Arbor, United States, associated to $^{67}$\\
\bigskip
% ...
\medskip
}
\end{flushleft}

\end{document}